\documentclass[
  journal=largetwo,
  manuscript=article-type,
  year=2020,
  volume=37,
]{cup-journal}

\usepackage{bm}

\usepackage{graphicx}

\usepackage{amsmath,amssymb}
\usepackage[nopatch]{microtype}
\usepackage{booktabs}
\usepackage{aas-macros}

\usepackage{hyperref}

\def\revision#1{{\bf\textcolor{blue}{{#1}}}}

\title{Can high-redshift AGN observed by JWST explain the EDGES absorption signal?}

\author{Alexandra Nelander}
\affiliation{School of Earth and Space exploration, Arizona State University, Tempe, AZ 85281, USA}
\email[Alexandra Nelander]{anelande@asu.edu}

\author{Christopher Cain}
\affiliation{School of Earth and Space exploration, Arizona State University, Tempe, AZ 85281, USA}

\author{Jordan C. J. D'Silva}
\affiliation{International Centre for Radio Astronomy Research (ICRAR), The University of Western Australia, M468, 35 Stirling Highway, Crawley, WA 6009, Australia}

\author{Peter H. Sims}
\affiliation{School of Earth and Space exploration, Arizona State University, Tempe, AZ 85281, USA}
\alsoaffiliation{Astrophysics Group, Cavendish Laboratory, J. J. Thomson Avenue, Cambridge CB3 0HE, UK}
\alsoaffiliation{Kavli Institute for Cosmology, Madingley Road, Cambridge CB3 0HA, UK}

\author{Rogier A. Windhorst}
\affiliation{School of Earth and Space exploration, Arizona State University, Tempe, AZ 85281, USA}

\author{Judd D. Bowman}
\affiliation{School of Earth and Space exploration, Arizona State University, Tempe, AZ 85281, USA}

\affiliation{Astrophysics Group, Cavendish Laboratory, J. J. Thomson Avenue, Cambridge CB3 0HE, UK}
\affiliation{Kavli Institute for Cosmology, Madingley Road, Cambridge CB3 0HA, UK}


\begin{document}

\begin{abstract}

The Experiment to Detect the Global Epoch of reionization 21 cm Signal (EDGES) has reported evidence for an absorption feature in the sky-averaged radio background near $78$ MHz.  A cosmological interpretation of this signal corresponds to absorption of 21 cm photons by neutral hydrogen at $z \sim 17$.  The large depth of the signal has been shown to require an excess radio background above the CMB and/or non-standard cooling processes in the IGM.  Here, we explore the plausibility of a scenario in which the EDGES signal is back-lit by an excess radio background sourced from a population of radio-loud AGN at high redshift.  These AGN could also explain the unexpected abundance of UV-bright objects observed at $z > 10$ by JWST.  We find that producing enough radio photons to explain the EDGES depth requires that nearly all high-$z$ UV-bright objects down to $M_{\rm UV} \gtrsim -15$ are radio-loud AGN and that the UV density of such objects declines by at most $1.5$ orders of magnitude between $z = 10$ and $20$.  In addition, the fraction of X-ray photons escaping these objects must be $\lesssim 1\%$ of their expected intrinsic production rate to prevent the absorption signal being washed out by early IGM pre-heating.  Re-producing the sharp boundaries of the absorption trough and its flat bottom require that the UV luminosity function, the fraction of UV light produced by AGN, and the X-ray escape fraction have fine-tuned redshift dependence.  We conclude that radio-loud AGN are an unlikely (although physically possible) candidate to explain EDGES because of the extreme physical properties required for them to do so.  
 
\end{abstract}

\section{Introduction}
\label{sec:intro}

The red-shifted 21 cm signal from neutral hydrogen lies at the frontier of modern cosmology.  During and prior to the epoch of reionization (EoR), neutral hydrogen in the intergalactic medium (IGM) absorbed or emitted 21 cm photons via its spin-flip transition, in principle allowing for direct observations of the neutral IGM~\citep{Madau1997,Shaver1999,Gnedin2004,Pritchard2008}.  Measurements of the sky-averaged signal and its fluctuations have been proposed as probes of reionization~\citep{Mesinger2007,Datta2007}, the formation and evolution of Pop III stars~\citep{Chuzhoy2006,Cruz2024}
and black holes~\citep{Ripamonti2008}, cosmological parameters~\citep{McQuinn2006,Bowman2007}, and alternative dark matter models~\citep{Valdes2007,Hibbard2022}, among others.  Numerous existing and forthcoming radio experiments are targeting one or more facets of this signal, including EDGES~\citep{Bowman2018}, SARAS~\citep{Singh2018}, the MWA~\citep{Pober2016}, HERA~\citep{Berkhout2024}, and the SKA~\citep{Koopmans2015}.  

The 21 cm signal between the formation of the first stars and the onset of the EoR (at $10 \lesssim z \lesssim 30$) is commonly referred to as the Cosmic Dawn (CD) signal.  Early efforts to detect it have targeted the sky-averaged signal, which is expected to be seen in absorption at these redshifts.  Recently, the EDGES experiment~\citep{Bowman2018} claimed evidence for the first detection of the CD signal - an absorption trough at $15 \lesssim z \lesssim 20$ reaching a differential brightness between the 21-cm spin temperature and radio background temperature of $-500$ mK at $z = 17$.  The amplitude of the signal is a factor of $\sim 2-3$ larger than the maximum amplitude predicted by concordance CD models~\citep[e.g.][]{Meiksin2020}, and the redshift evolution of the signal is also unusual~\citep{Mittal2022}.  Indeed, several works have questioned its cosmological origin.  Notably, the SARAS 3 experiment claimed a non-detection of the signal amplitude claimed by EDGES at $95\%$ confidence (\citet{Singh2022}, see also~\citet{Bevins2022}).  Concerns have also been raised over the techniques used to model the sky-averaged radio signal used by EDGES~\citep[e.g.][]{Hills2018,Sims2020,Cang2024}.  However, the possibility that the signal is of cosmological origin remains open and of interest.  

Several mechanisms have been proposed to explain the non-standard depth of the EDGES feature.  These generally involve either non-standard cooling of the neutral IGM, perhaps due to interactions with dark matter~\citep{Barkana2018,Datta2020,Halder2021}, or a radio background in excess of the CMB~(e.g. \citet{Feng2018b,Natwariya2021,Mittal2022b}, although see~\citet{Cang2024}).  In the latter scenario, which will be the focus of this work, several mechanisms have been proposed to produce the required excess radio emission.  These include accreting black holes~\citep{EwallWice2018,EwallWice2020}, an excess of faint galaxies at high redshifts~\citep{Mirocha2019}, or unexpected sources of galactic radio emission~\citep{Reis2020}.  The detection of an excess low-frequency radio background by ARCADE-2~\citep{Fixsen2011} at low redshift lends plausibility to this explanation (see also~\citet{Tompkins2023}).  

JWST observations are closing in on the redshift range probed by EDGES.  The latest constraints on the UV luminosity function (UVLF) of galaxies have pushed as high as $z = 14$~\citep{Adams2023,Donnan2024,Finkelstein2024}\footnote{See~\citet{PerezGonzalez2025,Castellano2025} for tentative constraints at higher redshifts.  }.  One key finding is an unexpectedly high abundance of bright galaxies at $z > 10$, which have strained pre-JWST models of galaxy formation~\citep{Chemerynska2024,Lu2025,Harikane2025,Whitler2025}, and perhaps even $\Lambda$CDM cosmology itself~\citep{Boylan-Kolchin2023}.  Explanations for this excess include increased star formation efficiency at high redshift~\citep{Jeong2025}, younger stellar populations~\citep{Donnan2025}, excess light from accreting black holes~\citep{Matteri2025}, and more exotic scenarios such as early dark energy~\citep{Shen2024}.  

A compelling possibility is that a large population of radio-loud accreting black holes at $10 \lesssim z \lesssim 20$ explains both of these puzzles.  There is some observational support for the possibility that a significant fraction of the $z > 10$ UV light seen by JWST is produced by AGN.  Recently,~\citet{Dsilva2023} fit the SEDs of bright galaxies at $z \sim 10.5$ and found that around half of the UV light could be explained by an AGN contribution (see also~\citet{Dsilva2025}).  Along the same lines,~\citet{Hegde2024} found that up to half the light produced by $z > 10$ objects could plausibly come from AGN without violating constraints on galaxy morphology~\citep{Ono2023}.  If a large fraction of these objects are radio-loud AGN that begin growing at $z > 15$, they may also produce enough radio emission at these redshifts to backlight EDGES in combination with the CMB~\citep{EwallWice2018,EwallWice2020}.  

This scenario finds additional motivation from observations of bright quasars at lower redshifts.  An increasingly large sample of bright AGN have been detected within the first billion years of cosmic history ($z \gtrsim 6$), with masses often inferred to be $10^9 M_{\odot}$ or more~\citep{Wu2015,Banados2018,Eilers2018,Yang2020a,Bennett2024}.  Indeed, such black holes are often observed to be over-massive relative to their host galaxies, suggesting a possible early onset of AGN growth relative to stellar formation~\citep{Sun2025}.  Growing such massive black holes by these redshifts generally requires massive seeds~\citep{Woods2019}, super-Eddington accretion~\citep{Johnson2022} and/or low radiative efficiencies~\citep{Morey2021}.  In any case, it is reasonable to expect that rapidly growing AGN should exist at much higher redshifts.  

In this work, we investigate the possibility that a large population of $z > 10$ AGN could explain JWST observations and the EDGES signal.  We use observations of the UVLF up to $z = 14$ from JWST, extrapolated to higher redshifts, to model the excess radio background produced by AGN.  We also account for the effect AGN would have on Ly$\alpha$ coupling of the 21 cm spin temperature to the IGM gas temperature and the pre-heating of the IGM by X-rays.  This work is organized as follows.  In \S\ref{sec:modeling}, we describe our formalism for modeling the sky-averaged 21 cm signal.  \S\ref{sec:results} presents our main results, and in \S\ref{sec:implications} we discuss implications of our findings for interpreting high-redshift observations.  We conclude in \S\ref{sec:conc}.  Throughout, we assume the following cosmological parameters: $\Omega_m = 0.305$, $\Omega_{\Lambda} = 1 - \Omega_m$, $\Omega_b = 0.048$, $h = 0.68$, $n_s = 0.9667$ and $\sigma_8 = 0.82$, consistent with~\citet{Planck2018} results. Distances are in co-moving units unless otherwise specified. 

\section{Modeling the cosmic dawn global 21 cm signal}
\label{sec:modeling}

\subsection{Basics of the signal}
\label{subsec:basics}

The sky-averaged 21 cm absorption temperature during the CD and reionization is given by
\begin{equation}
    \label{eq:T21}
    T_{21}(z) = \bar{T}_{21}^0 (1+z)^{\frac{1}{2}} x_{\rm HI}(z) \left[1 - \frac{T_{\rm radio}(z)}{T_{\rm S}(z)}\right]
\end{equation}
where $\bar{T}_{21}^0 = 8.5 \left(\frac{\Omega_b}{0.044}\right)\left(\frac{h}{0.7}\right)\left(\frac{\Omega_m}{0.27}\right)^{\frac{1}{2}}$ mK is independent of redshift~\citep{Giri2019b}, $x_{\rm HI}(z)$ is the mass-averaged IGM neutral fraction, $T_{\rm radio}$ is the radio background temperature at $1.4$GHz and $T_{\rm S}$ is the spin temperature of neutral IGM gas.  During the CD, $x_{\rm HI} = 1$, and the redshift evolution of the signal is driven by that of $T_{\rm radio}$ and $T_{\rm S}$.  In standard scenarios, the radio background is taken to be that of the CMB - namely, $T_{\rm radio} = T_{\rm CMB} = 2.7 (1+z)$ K.  In cases with an external radio background $T_{\rm ex}$, we write $T_{\rm radio} = T_{\rm CMB} + T_{\rm ex}$.  Note that $T_{\rm ex} > 0$ is required, in canonical 21 cm models, to achieve $T_{21}$ much less than $-200$ mK by $z = 17$ (see e.g. Figure 2 of~\citet{Cang2024}).

When the first stars form, the UV radiation they produce couples $T_{\rm S}$ to the gas kinetic temperature via Ly$\alpha$ coupling.  Ignoring collisional coupling (which is unimportant at $z < 30$), the coupling is described by
\begin{equation}
    \label{eq:Ts}
    T_{\rm S}^{-1} = \frac{T_{\rm radio}^{-1} + x_{\alpha} T_{\rm K}^{-1}}{1 + x_{\alpha}}
\end{equation}
where $x_{\alpha}$ is the Ly$\alpha$ coupling coefficient.  In canonical 21 cm models (e.g.~\citet{Furlanetto2006c}), coupling is started by pop III stars, since these are the first sources of UV radiation.  Prior to pre-heating by the first X-ray sources, the IGM has kinetic temperature $T_{\rm K} < T_{\rm CMB}$, such that the average signal is negative (appears in absorption, Equation ~\ref{eq:T21}).  Heating of the neutral IGM by X-rays before reionization (``pre-heating'') likely raises $T_{\rm K}$ well above $T_{\rm CMB}$ by $z \sim 10$, such that the EoR signal appears in emission.  Later in this section, we will describe how we model the key components driving the CD signal - the excess radio background $T_{\rm ex}$ (\S\ref{subsec:radio_AGN}), Ly$\alpha$ coupling (\S\ref{subsec:lya_coupling}), and X-Ray heating (\S\ref{subsec:xray_heating}).  

\subsection{JWST UV Luminosity Functions}
\label{subsec:UVLF}

\begin{figure*}[hbt!]
    \centering
    \includegraphics[scale=0.72]{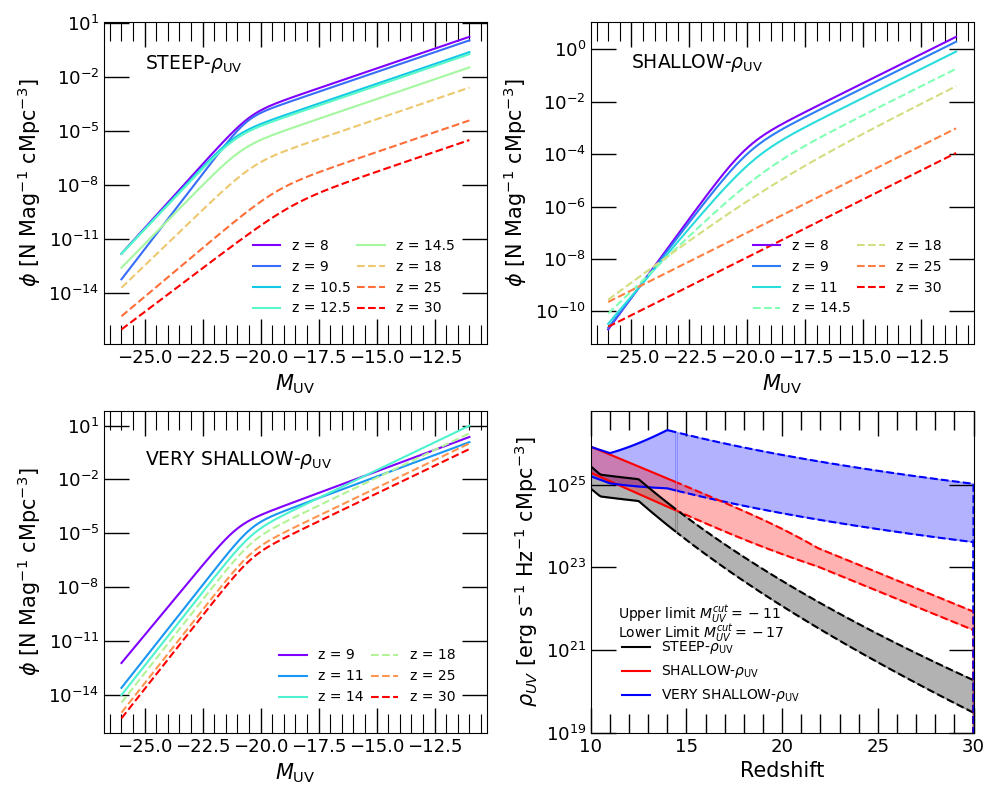} 
    \caption{UVLFs assumed in this work and their associated UV luminosity densities.  {\bf Top Left:} UVLFs for the \textsc{steep-$\rho_{\rm UV}$} model, which uses measurements from~\citet{Adams2023} at $8 \leq z \leq 12.5$ and~\citet{Donnan2024} at $z = 14.5$.  Solid curves denote redshifts where measurements are available, and dashed curves are extrapolations up to $z = 30$.  {\bf Top Right:} the same for our \textsc{shallow-$\rho_{\rm UV}$} model, which extrapolates the redshift-dependent fits to the UVLF parameters in Equations 3-6 of~\citet{Donnan2024}.  {\bf Bottom Left:} our \textsc{very shallow-$\rho_{\rm UV}$} model, based on extrapolating the UVLF measurements at $z = 9$, $11$, and $14$ from~\citet{Finkelstein2024}.  {\bf Bottom Right:} the UV luminosity density, $\rho_{\rm UV}$, for all three models.  The shaded range denotes the difference between assuming bright ($M_{\rm UV} < -17$) and faint ($M_{\rm UV} < -11$) integration limits.  Dashed lines denote extrapolations from measurements.  }
    \label{fig:UVLF}
\end{figure*}

In this work, we base our model for the CD 21 cm signal on measurements of the UV luminosity functions (UVLF) by JWST up to $z \approx 14$~\citep[e.g.][]{Adams2023,Finkelstein2024,Donnan2024}.  Our premise is that a significant fraction of UV-luminous objects during the CD are radio-loud AGN.  In Figure~\ref{fig:UVLF}, we show three models for the UVLF considered in this work, which are based on recent JWST observations at $10 < z < 14$.  The top left, top right, and bottom left panels show the UVLF for each model, with solid curves denoting redshifts where measurements are available and dashed lines denoting extrapolations to higher redshifts (discussed below).  In the bottom right, we show the UV luminosity density, $\rho_{\rm UV}$, for each model, with dashed lines indicating extrapolations.  The shaded region shows the range obtained from integrating each set of UVLFs down to limiting magnitudes of $M_{\rm UV} = -17$ and $-11$.  

The top left panel shows our \textsc{steep-$\rho_{\rm UV}$} model, which uses measurements at $8 \leq z \leq 12.5$ from~\citet{Adams2023} and the~\citet{Donnan2024} measurement at $z = 14.5$.  These fits (and all the others we use) assume a double power-law for the UVLF of the form
\begin{gather*}
   \frac{dn}{dM_{\rm UV}} = \frac{\phi^{\ast}\ln 10}{2.5}\left(\left[ 10^{0.4(M_{\rm UV} - M_{\ast})} \right]^{\alpha+1}
   + \left[ 10^{0.4(M_{\rm UV} - M_{\ast})} \right]^{\beta+1}  \right)
\end{gather*}
where $\alpha$ is the faint-end slope, $\beta$ is the bright end slope, $M_{\ast}$ is the cutoff between bright and faint ends, and $\phi^\ast$ is the amplitude parameter.  We linearly extrapolate the shape parameters in redshift - $\alpha$, $\beta$, and $M_{\ast}$ - between the $z = 12.5$ and $14.5$ measurements, and do a log-linear extrapolation of  $\phi^\ast$.  This UVLF model falls off rapidly with redshift, with $\rho_{\rm UV}$ dropping $4$ orders of magnitude between $z = 14$ and $30$ (bottom right). 
The top right panel shows our \textsc{shallow-$\rho_{\rm UV}$} model, which uses Equations 3-6 for the UVLF parameters in~\citet{Donnan2024} to predict the UVLF out to $z = 30$.  These equations were calibrated to best match a collection of measurements at $8 < z < 14$ from~\citep{Bowler2016,Bowler2020,Donnan2023,McLeod2023}.  They predict that the bright-end slope, $\beta$, should increase rapidly with redshift, becoming larger than the faint-end slope, $\alpha$, at $z \sim 25$.  To avoid un-physical behavior in the UVLF,\footnote{The symmetric form of the double power law function used to fit the UVLF implies that $\max(\alpha,\beta)$ behaves as the faint-end slope.  Thus, allowing $\beta > \alpha$ swaps the function of the two variables.  Our condition avoids this behavior.  } we impose the condition that $\beta \leq \alpha$ at all redshifts.  

The bottom left panel shows our \textsc{very shallow-$\rho_{\rm UV}$} model, which is based on recent measurements at $z = 9$, $11$, and $14$ by~\citet{Finkelstein2024}.  They find that $\rho_{\rm UV}$ integrated down to $M_{\rm UV} = -17$ is nearly flat (evolves only a factor of a few) between redshifts $9$ and $14$.  They further find a relatively steep faint-end slope of $\alpha = -2.55$ at $z = 14$, which predicts an increase in $\rho_{\rm UV}$ with redshift when integrated to $M_{\rm UV} = -11$ (see lower right panel).  Extrapolating their measurements to $z = 30$ (assuming $\alpha = -2.55$ at $z > 14$) predicts only one order of magnitude decrease in $\rho_{\rm UV}$ across this redshift range.  For this model, we must impose the additional condition that $\phi^\ast(z > 30) = 0$ to avoid un-physically high UV light production extending into the cosmic dark ages (before there was sufficient structure formation to produce halos capable of hosting galaxies).  

Estimating the radio and X-ray emissions of an AGN from its UV luminosity $L_{\rm UV}$ requires an estimate of its black hole mass, $M_{\rm bh}$.  The associated bolometric luminosity is given by~\citep{Hegde2024}
\begin{equation}
    \label{eq:Lbh}
    L_{\text{bh}} = 3.3 \times 10^4 \times \eta_{\text{edd}} M_{\text{bh}} 
\end{equation}
where $\eta_{\rm edd}$ is the Eddington ratio.  Following~\citet{Hegde2024}, we map bolometric luminosity to UV luminosity using the procedure described in~\citet{Marconi2004}, which yields
\begin{equation}
    \label{eq:LUV}
    L_{\rm UV} = \frac{ L_{\text{bh}}} {\nu_b 10^{0.8 - 0.067\gamma + 0.017\gamma^2 - 0.0023\gamma^3}} 
\times \left( \frac{\nu_{\text{UV}}}{\nu_{\text{b}}} \right)^{-\alpha_{\rm UV}} 
\end{equation}
where $\nu_{\rm b} = 6.67 \times 10^{14}$ Hz and $\nu_{\rm UV} = 2 \times 10^{15}$ Hz and $\gamma = \log_{10} (L_{\text{bh}}/L_{\odot}) - 12 $.   We assume a UV slope of $\alpha_{\rm UV} = 0.6$, typical for bright quasars at lower redshifts~\citep{Lusso2015}.  Assuming a given object's UV emission is dominated by the AGN, Equation ~\ref{eq:Lbh}-\ref{eq:LUV} can be used to map between $L_{\rm UV}$ and $M_{\rm bh}$.  We have tested how much our results vary across the range $0.45 < \alpha_{\rm UV} < 0.7$~\citep{Berk2001,Telfer2002} and found a difference in $T_{21}$ of $<10\%$ across this range. This effect is sub-dominant relative to other sources of uncertainty in our analysis.  

\subsection{Excess radio background from AGN}
\label{subsec:radio_AGN}

The total radio emissivity of the AGN population at redshift $z$ and frequency $\nu$ can be written as
\begin{equation}
    \epsilon(\nu,z) = f_{\rm l} f_{\rm bh}\int_{-\infty}^{M_{\rm UV}^{\rm cut}}dM_{\rm UV} \frac{dn}{dM_{\rm UV}}L_{\nu,{\rm radio}}(M_{\rm UV})
\label{eq:epsilon_radio}
\end{equation}
where $f_{\rm l}$ is the fraction of AGN that are radio-loud, $M_{\rm UV}^{\rm cut}$ is the cutoff between objects dominated by AGN and stellar light, $f_{\rm bh}$ is the fraction of objects with $M_{\rm UV} < M_{\rm UV}^{\rm cut}$ that are AGN, $dn/dM_{\rm UV}$ is the UVLF, and $L_{\nu,{\rm radio}}(M_{\rm UV})$ is the radio luminosity of an AGN with magnitude $M_{\rm UV}$.  Note that $M_{\rm UV}^{\rm cut}$ is defined here as the cutoff in the UVLF between AGN and stellar-driven UV sources, {\it not the faint-end turnover of the UVLF}.  We estimate the radio luminosity at $5$GHz, $L_{\rm 5GHz}$, using recent measurements of the fundamental plane of radio-loud quasars from~\citet{Bariuan2022}, 
\begin{gather}
    \log_{10} [L_{5 \rm GHz}/{\rm erg \text{ } s^{-1}}] = \nonumber \\ b + \xi_{x} \log_{10} [L_{2-10}/{\rm erg \text{ } s^{-1}}] + \xi_m \log_{10} [M_{\rm bh}/M_{\odot}]
    \label{eq:L5}
\end{gather}
where $\xi_x = 1.12$, $\xi_m = -0.210$, and $b = -5.64$ and $L_{2-10}$ is the intrinsic X-ray luminosity of the AGN between $2$ and $10$ keV.  This is given by
\begin{equation}
\label{eq:L_2_10}
L_{2-10} = k_{\rm bol}^{-1} \eta_{\rm edd} L_E
\end{equation}
where $k_{\rm bol}^{-1} = 0.07$ is the bolometric correction factor~\citep{Lusso2010} and $L_E = 1.26 \times 10^{38} $ erg s$^{-1}$ $\frac{M_{\rm bh}}{M_{\odot}}$ is the Eddington luminosity. 
The radio luminosity per unit frequency $L_{\nu, {\rm radio}}$ is then given by
\begin{equation}
    L_{\nu,{\rm rad}} = \frac{L_{5 \rm GHz}}{5 \text{GHz}}\left(\frac{1.4 {\rm GHz}}{5 {\rm GHz}}\right)^{-0.5}\left(\frac{\nu}{1.4 {\rm GHz}}\right)^{-\alpha_{\rm radio}}
\end{equation}
where the first factor corrects for the re-scaling from $1.4$GHz to $5$GHz assumed\footnote{Measurements used in~\citet{Bariuan2022} are made close to $1.4$GHz and extrapolated to $5$GHz assuming $\alpha_{\rm radio} = 0.5$, so this re-scaling recovers the exact measured flux at $1.4$GHz.  Physically, there is significant scatter in $\alpha_{\rm radio}$ for quasars of different types~\citep{Gloudemans2022} - however, our results are sensitive only to radio emission at frequencies close to $1.4$GHz, and thus modest differences in $\alpha_{\rm radio}$ have a minimal effect on our results.  }. in~\citet{Bariuan2022}, and we take $\alpha_{\rm radio} = 0.5$ following the assumed scaling in~\citet{Bariuan2022}, although our results are reasonably insensitive to the assumed frequency dependence.  The corresponding radio temperature at $1.4$ GHz ($T_{\rm ex}$ in \S\ref{subsec:basics}) is given in terms of $\epsilon(\nu,z)$ by Equation 5-6 of~\citet{EwallWice2018}.  The fundamental plane relation in~\citet{Bariuan2022} reports a logarithmic scatter of $\sigma_{\rm R} \approx 0.5$ (in a fixed redshift bin).  Assuming that the underlying distribution is lognormal around the best-fit relation, this corresponds to a factor of $\approx 2$ boost for the average radio emission relative to the best-fit relation in log space.  We multiply the $L_{\rm 5GHz}$ given by Equation ~\ref{eq:L5} by a factor of $2$ to correct for intrinsic scatter.  

There is considerable uncertainty in the radio-loud fundamental plane given by Equation ~\ref{eq:L5}.  Figure 4 of~\citet{Bariuan2022} shows that there is nearly an order of magnitude of scatter on either side of their best-fit relation, and they noted potential redshift evolution.  Furthermore,~\citet{Wang2024} found somewhat lower radio emission than in~\citet{Bariuan2022}, and commented on possible extra dependencies not considered in that work (see also~\citet{Qing-Chen2025}).  There is also variation (a dex or more) in the hard X-ray bolometric correction factors observed for AGN.  Our choice of $k_{\rm bol} = 0.07^{-1} = 14.3$ is on the low end of the distribution observed in~\citet{Lusso2010,Lusso2012}, and in more recent work by~\citet{Spinoglio2024}.  It is also consistent with the values predicted for low-luminosity AGN by~\citet{Duras2020}.  Higher $k_{\rm bol}$ would result in lower $L_{2-10}$ at fixed $M_{\rm bh}$, and thus reduced $L_{\rm 5GHz}$, which would in turn produce a weaker radio background.  As we will see in later sections, assuming a relatively low $k_{\rm bol}$ is conservative for the purposes of this work.

\subsection{Ly $\alpha$ coupling}
\label{subsec:lya_coupling}

UV photons with energies between $10.2$ and $13.6$ eV from the first stars/AGN coupled the spin temperature of the 21 cm transition to the kinetic temperature of the neutral gas, a process known as Ly$\alpha$ coupling.  This occurs by means of the Wouthuysen-field effect via transitions to the 2p state of neutral hydrogen~\citep{Wouthuysen1952,Field1958,Hirata2005}.  
We describe in this section how we model the contribution to Ly$\alpha$ coupling from pop II stars/AGN and pop III stars.  

\subsubsection{Pop II stars \& AGN}
\label{subsubsection:Lya_AGN}

The Ly$\alpha$ coupling coefficient, $x_{\alpha}$ (see Equation ~\ref{eq:Ts}) is given by

\begin{equation}
\label{eq:xalpha}
x_\alpha = \frac{4 \pi e^2 f_\alpha T^\ast}{27 A_{10} T_{\text{CMB}} m_e}S_\alpha n_\alpha
\end{equation}
where $T_{\ast} = 68.2 \, \text{mK}$, \( f_{\alpha} = 0.4162 \), \( A_{10} = 2.87 \times 10^{-15} \, \text{s}^{-1} \), $e$ is the electron charge, $T_{\rm CMB}$ is the CMB temperature and $m_e$ is the electron mass.  The scattering correction, $S_{\alpha}$, is given by

\begin{equation}
\label{eq:salpha}
S_\alpha = \exp\left( -0.013 \tau^{1/3}_{\rm GP} / T_{\rm K}^{2/3} \right)
\end{equation}

where $\tau_{\rm GP}$ is the Gunn-Petersen optical depth and $T_{\rm K}$ is the gas kinetic temperature. The number density of Ly$\alpha$ photons is given by~\citep{Madau2018},

\begin{equation}
\label{eq:nalpha}
n_\alpha(z) = (1 + z)^2 \sum_{n=2}^{\infty} P_{np} \int_{z}^{z_{\text{max}}(n)} \dot{n}_{\nu'}(z') \frac{dz'}{ H(z')}
\end{equation}

where the sum runs over the transitions from $n$th excited state to the ground state and $P_{\rm np}$ is the probability for an HI atom in the np excited state to produce a Ly$\alpha$ photon.  Here, $\nu' = \nu_{\rm n}(1+z')/(1+z)$, where $\nu_{\rm n}$ is the frequency of the $n$th excited state.  For the $n$th excited state, the upper limit of the integral is

\begin{equation}
\label{eq:zmax}
z_{\text{max}}(n) = \left( 1+z \right) 
\frac{1- \frac{1}{n^2}}{1-\frac{1}{(n+1)^2}}
\end{equation}

The number density of UV photons produced by pop II stars and AGN is given in terms of the UV luminosity density as
\begin{equation}
    \label{eq:n_nu_z}
    \dot{n}_\nu(z) = \left(\frac{\rho_{\rm UV}(z)}{h_{\rm p} \nu_{\rm UV}}\right) \left(\frac{\nu}{\nu_{\rm UV}}\right)^{-\alpha_{\rm UV}-1}
\end{equation}
where again we have taken $\alpha_{\rm UV} = 0.6$. We assume here that the faint-end turnover happens at $M_{\rm UV} = -11$, such that we integrate down to that limit to get $\rho_{\rm UV}$ in Equation ~\ref{eq:n_nu_z}.  In our model, the UVLF at $M_{\rm UV} < M_{\rm UV}^{\rm cut}$ is dominated by AGN, and fainter objects are dominated by pop II stars.

\subsubsection{Pop III Stars}
\label{subsubsec:popIII}

Pop III stars were probably the first drivers of Ly$\alpha$ coupling.  These first stars likely formed predominantly in halos well below the atomic cooling limit (masses $< 10^8 M_{\odot}$), the so-called ``mini-halos''~\citep{MiraldaEscude2003}.  As such, these stars likely formed in halos much below the visibility limits of current observations~\citep{Windhorst2018}, and/or may not contribute significantly to the total star formation rate at $z < 15$~\citep{Visbal2020}. As such, that they would likely not be accounted for in our extrapolations of the UVLF in Figure~\ref{fig:UVLF} (although see~\citet[e.g.][]{Fujimoto2025,Venditti2025} for suggestions that some Pop III stars may live in more massive halos).  
We include the contribution of Pop III stars to Ly$\alpha$ coupling using the analytic prescription described in~\citet{McQuinn2012}.  The star formation rate density of pop III stars is given by

\begin{equation}
    \label{eq:rhoSFRIII}
    \rho_{\rm SFR}^{\rm III} = f_\ast \rho_{b} \frac{d}{d t} \left[\rho_{\rm m}^{-1}\int_{M_{\rm crit}}^{M_{\rm atom}} dM \left(\frac{dn}{dM}\right) M\right]
\end{equation}

where $f_{\ast}$ is the pop III star formation efficiency, $\rho_{\rm b}$ and $\rho_{\rm m}$ are the baryon and total matter densities, and $dn/dM$ is the Sheth-Torman halo mass function~\citep{Sheth1999}.  The expression in brackets is the fraction of mass collapsed in molecular cooling halos hosting pop III star formation.  The upper limit of the integral is the atomic cooling limit, which we take to be $M_{\rm atom} = 10^{8} M_{\odot}$, and the lower limit is the critical mass for molecular hydrogen formation~\citep{Machacek2001}, 
\begin{equation}
\label{eq:Mcrit}
M_{\rm crit} = 2.5\times 10^5 \, + \, 8.7\times10^5 F^{0.47}_{LW,21} \, M_{\odot}
\end{equation}
where the normalized Lyman-Werner intensity is
\begin{equation}
    \label{eq:FLW21}
    F_{LW,21} = 20 \, \left(\frac{1+z}{21}\right) \, e^{-\tau_{LW}} \text{erg s}^{-1}
\end{equation}
and we take $\tau_{\rm LW} = 1.5$.  We can approximate the coupling coefficient due to UV emission from pop III star by inverting Equation 7 of~\citet{McQuinn2012}, which yields
\begin{equation}
    \label{eq:xalphaIII}
    x_{\alpha}^{\rm III} = \frac{\rho_{\rm SFR}^{\rm III}}{1.7\times10^{-3} M_{\odot} {\rm yr}^{-1}{\rm cMpc}^{-3}} \frac{N_{\alpha}}{10^4}\left(\frac{1+z}{21}\right)
\end{equation}
and the total coupling coefficient is given by linearly adding this term to the pop II/AGN component given by Equation ~\ref{eq:xalpha}.  

\begin{figure}[hbt!]
    \includegraphics[scale=0.49]{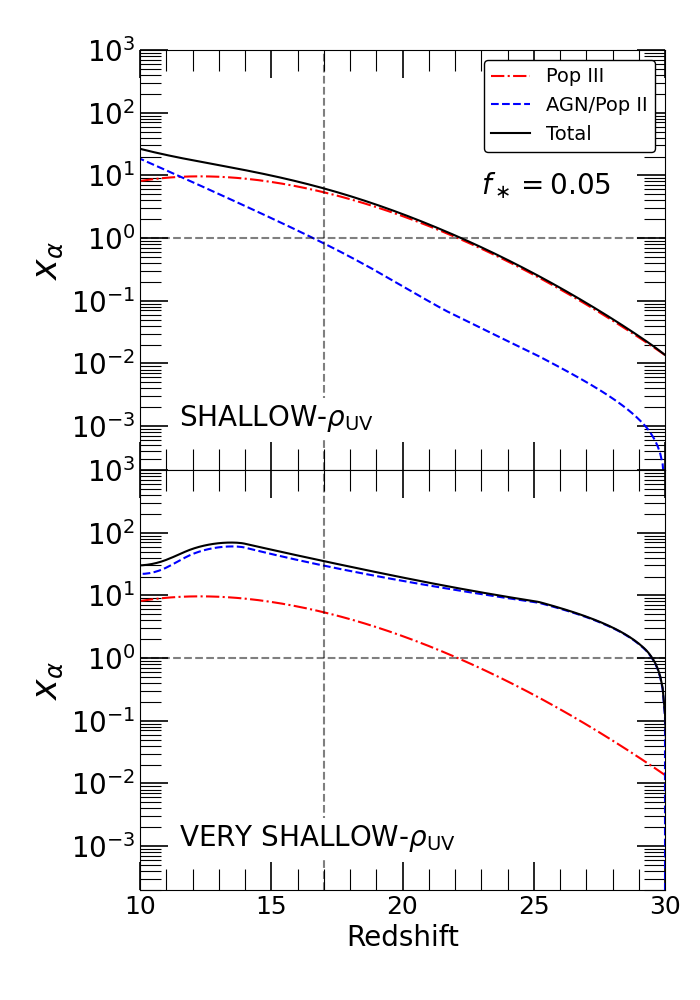} 
    \caption{Ly$\alpha$ coupling coefficient in our \textsc{shallow-$\rho_{\rm UV}$} (top) and \textsc{very shallow-$\rho_{\rm UV}$} (bottom) models.  The solid black line denotes the total $x_{\alpha}$, and the dot-dashed red and dashed blue lines the contributions from pop III stars and pop II star/AGN, respectively.  The vertical dashed line denotes $z = 17$, the central EDGES redshift, and the horizontal line  $x_{\alpha} = 1$, at which coupling is halfway complete.  We choose $f_{\ast} = 0.05$ for pop III stars, such that coupling is well underway by $z = 20$, as required by EDGES.  In the \textsc{shallow-$\rho_{\rm UV}$} case, coupling is dominated by pop III stars, but in the \textsc{very shallow-$\rho_{\rm UV}$} case it is dominated by the pop II/AGN component.  }
    \label{fig:xalpha}
\end{figure}

In Figure~\ref{fig:xalpha}, we show the evolution of $x_{\alpha}$ with redshift in our \textsc{shallow-$\rho_{\rm UV}$} (top) and \textsc{very shallow-$\rho_{\rm UV}$} (bottom) models.  In each panel, the solid black curve shows the total $x_{\alpha}$, while the dot-dashed red and dashed blue curves show the contribution from pop III stars and pop II stars/AGN, respectively.  The vertical dashed line denotes $z = 17$, the central redshift of the EDGES signal, and the horizontal line indicates $x_{\alpha} = 1$, for which coupling is halfway complete.  We chose $f_\ast = 0.05$ (within the range assumed in~\citet{McQuinn2012}) so that coupling is well underway by $z = 20$, as required by the EDGES signal.   In the \textsc{steep-$\rho_{\rm UV}$} (not shown) and \textsc{shallow-$\rho_{\rm UV}$}, pop III stars dominate Ly$\alpha$ coupling at $z > 15$ for this value of $f_\ast$.  However, in the \textsc{very shallow-$\rho_{\rm UV}$} model, coupling occurs abruptly at $z = 30$ and is dominated by the pop II/AGN component\footnote{Indeed, coupling in this model would begin much earlier than $z = 30$ without our condition that $\phi^\ast(z > 30) = 0$.  }.  

We note that a number of works more recent than~\citet{McQuinn2012} have provided updated estimates of $M_{\rm crit}$ (Equation~\ref{eq:Mcrit}), which take into account more accurately the effects of Lyman Werner suppression and baryon-dark matter streaming velocities~\citep[e.g.][]{Kulkarni2021,Nebrin2023,Hegde2023}.  These works, in general, find somewhat lower $M_{\rm crit}$ than we assume here, which would result in earlier, faster Ly$\alpha$ coupling. However, $M_{\rm crit}$ is mostly degenerate with $f_\ast$, which we tuned to produce coupling within a redshift range favorable for EDGES.  As such, using these improved estimates of $M_{\rm crit}$ would have little impact on our results in this work.

\subsection{X-Ray heating}
\label{subsec:xray_heating}

X-ray photons have ionizing cross-sections small enough to allow them to travel cosmological distances through the neutral IGM.  The first X-ray sources are believed to have pre-heated the IGM to $T_{\rm K} > T_{\rm CMB}$ before the onset of reionization~\citep{Fialkov2014c,HERA2023}.  This resulted in $T_{\rm radio}/T_{\rm S} \rightarrow 0$ in Equation ~\ref{eq:T21}, driving the signal out of absorption and into emission.  The steep rise of the EDGES absorption signal towards $0$ at $z \approx 15$ is most naturally interpreted as the result of X-ray heating.  

Absent non-standard heating or cooling channels, the temperature of the neutral IGM before reionization is given by
\begin{equation}
\label{eq:dTdt}
\frac{dT}{dt} = -2H(z)T + \frac{2}{3 n_{\text{tot}} k_b} H_{\text{X-ray}}
\end{equation}
where $H_{\rm X-ray}$ is the heating rate from X-rays, $n_{\rm tot}$ is the particle number density of the IGM, and the other symbols have their usual meanings.  In this work, we assume that all X-ray heating is sourced by AGN, which gives a conservative estimate of the heating rate.  In reality, additional contributions are expected from pop III stars~\citep{Xu2016} and/or pop II high-mass X-ray binaries~\citep[HMXBs,][]{Fialkov2014c,Madau2017}.  

The X-ray luminosity at frequency $\nu$ by an AGN with magnitude $M_{\rm UV}$ is given by
\begin{equation}
\label{eq:Lx}
L_X(\nu,M_{\rm UV}) = \left(\frac{\nu}{\nu_2}\right)^{-\alpha_{\rm X}}
\frac{(1-\alpha_{\rm X}) L_{2-10}(M_{\rm UV})}{\nu_2 \, [5^{(1-\alpha_{\rm X})}-1]}
\end{equation}
where $\nu_{2} \equiv 2$ keV/$h_{\rm p}$ and the factor of $5$ is the ratio of $10$ keV and $2$ keV, $L_{2-10}$ is given by Equation ~\ref{eq:L_2_10}, and $\alpha_{\rm X}$ is the slope of the X-ray spectrum of AGN.  Following~\citet{EwallWice2018}, we assume $\alpha_{\rm X} = 0.9$.  We have denoted in Equation ~\ref{eq:Lx} that $L_{2-10}$ is a function of $M_{\rm UV}$ (Equation ~\ref{eq:Lbh}-\ref{eq:LUV}).  The co-moving emissivity of X-ray photons at frequency $\nu$ and redshift $z$ is given by

\begin{equation}
    \label{eq:NdotX}
    \dot{N}_{\rm X}(z,\nu) = f_{\rm esc}^{\rm X} f_{\rm bh} \frac{1}{h_{\rm p}\nu} \int_{-\infty}^{M_{\rm UV}^{\rm cut}} dM_{\rm UV} \frac{dn}{dM_{\rm UV}} L_{\rm X}(M_{\rm UV},\nu)
\end{equation}
where $f_{\rm esc}^{\rm X}$ is the escape fraction of X-ray photons, which we assume (for simplicity) is independent of photon energy.  The number density of X-Ray photons in the universe at frequency $\nu$ and redshift $z$ is given by

\begin{equation}
\label{eq:NX}
N_{\rm X}(\nu, z) = \int_{\infty}^{z} dz' \left|\frac{dt'}{dz'}\right| \dot{N}_{\rm X}(z',\nu') \exp[-\tau'(\nu',z',\nu,z)]
\end{equation}
where $\nu' = (1+z')/(1+z)\nu$ and $\tau'$ is the optical depth encountered by X-rays emitted at $z'$ with frequency $\nu'$ by the time they redshift to frequency $\nu$ at redshift $z$.  We can approximate this by
\begin{equation}
    \label{eq:tau_nu}
    \tau'(\nu',z',\nu,z) \approx \kappa(\nu,z) c (t - t')
\end{equation}
where $t$ and $t'$ are the cosmic times at redshifts $z$ and $z'$ (respectively), and $\kappa(z,\nu)$ is the absorption coefficient of the IGM to X-rays, given by
\begin{equation}
    \label{eq:kappa_nu}
    \kappa(z,\nu) = \sigma_{\rm HI}(\nu) n_{\rm H}(z) + \sigma_{\rm HeI}(\nu) n_{\rm He}(z)
\end{equation}
where $\sigma$ and $n$ are the absorption cross-sections and number densities for HI and HeI in a neutral IGM.  Note that Equation ~\ref{eq:tau_nu} neglects the dependence of $\kappa$ on $z'$ and $\nu'$ along the line of sight.  Fortunately, this approximation is reasonable thanks to the frequency dependence of the cross-sections of HI and HeI\footnote{This fortuitous cancellation occurs because the HI cross-section, which dominates the absorption rate, scales like $\nu'^{-2.75} \propto (1+z')^{-2.75}$, which almost exactly cancels the $(1+z')^3$ scaling of $n_{\rm HI}$ in a neutral IGM.  }.  The heating rate is then given by
\begin{equation}
\label{eq:HXray}
H_{\text{X-ray}}(z) = f_h n_{\rm H}(z) (1+z)^3 \int_{\nu_{0.5}}^{\nu_{10}} N_{\rm X}(\nu,z) \, c \, \sigma_{\text{HI}}(\nu) \, (h_{\rm p} \nu) \, d\nu 
\end{equation}
where $f_h$ is the fraction of injected energy that contributes to heating the IGM and we have neglected the sub-dominate contribution to the heating rate from ionizations of He. Following~\citet{Furlanetto2006}, we assume $f_h = 0.2$.  We have integrated down to $0.5$ keV to include heating from soft X-rays. 

We note here that our analysis neglects the effects of heating from Ly$\alpha$ photons and the CMB.  \cite{Reis2021} found that, in cases where heating from other sources (such as X-rays, see below) is very low, it can have a significant effect on the maximum depth of the 21 cm absorption feature.  As we will see in the next section, the signal is hard to reproduce with radio emission from AGN even under optimistic circumstances.  Since our goal here is to understand the minimal requirements for our model to achieve the EDGES depth at $15 < z < 20$, neglecting this effect is a conservative assumption with respect to our main results.  At $z < 15$, \cite{Reis2021} found that this heating source negligibly affects $T_{21}$ if the gas temperature is $\gtrsim 100$ K by $z = 10$, which it is in all the models we consider below that include heating effects.  As such, we expect that our results would be qualitatively unaffected by the inclusion of Ly$\alpha$ and CMB heating.  

\begin{figure}
    \centering
    \includegraphics[scale=0.46]{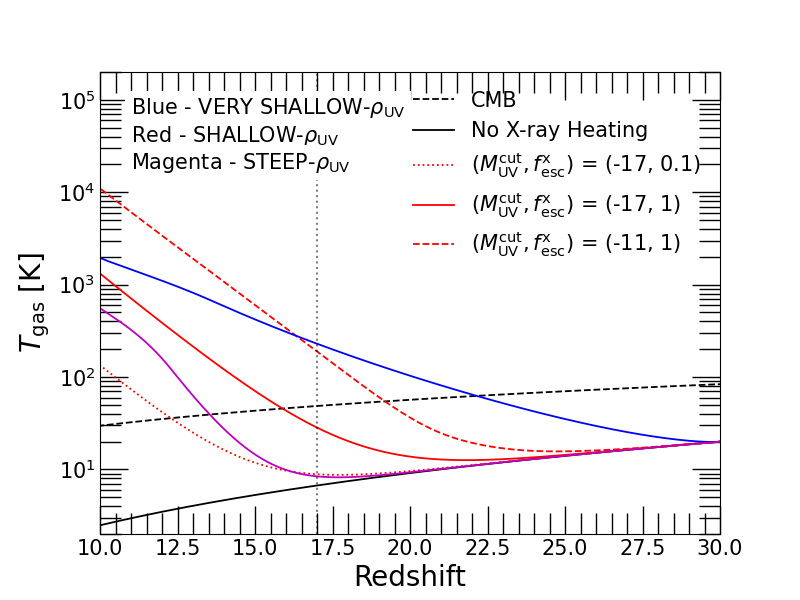} 
    \caption{Examples of thermal histories predicted by our AGN-driven pre-heating model for several representative scenarios.  The solid and dashed black curves show the temperature absent pre-heating (the adiabatic limit) and the CMB temperature, respectively.  For the \textsc{shallow-$\rho_{\rm UV}$} model, we show $T_{\rm K}$ for $(f_{\rm esc}^{\rm X},M_{\rm UV}^{\rm cut}) = (0.1,-17)$, $(1,-17)$, and $(1,-11)$ as the red dotted, solid, and dashed curves, respectively.  We also show the \textsc{steep-$\rho_{\rm UV}$} and \textsc{very shallow-$\rho_{\rm UV}$} models assuming $(f_{\rm esc}^{\rm X},M_{\rm UV}^{\rm cut}) = (1,-17)$ as the magenta and blue solid curves.  Models with higher $f_{\rm esc}^{\rm X}$, fainter $M_{\rm UV}^{\rm cut}$, and higher UV densities predict earlier pre-heating and higher temperatures.  }
    \label{fig:gas_temp}
\end{figure}

In Figure~\ref{fig:gas_temp}, we show the evolution of the gas temperature predicted by our AGN heating model for several scenarios.  The black solid curve is ``adiabatic limit'', without any pre-heating, and the dashed black curve shows $T_{\rm CMB}$.  The red dotted, solid, and dashed curves show $T_{\rm K}$ vs. $z$ for our \textsc{shallow-$\rho_{\rm UV}$} model assuming several combinations of $f_{\rm esc}^{\rm X}$ and $M_{\rm UV}^{\rm cut}$, indicated in the legend, in order of increasing heating rate.  The solid magenta and blue curves show $T_{\rm K}$ for $f_{\rm esc}^{\rm X} = 1$ and $M_{\rm UV}^{\rm cut} = -17$ for the \textsc{steep-$\rho_{\rm UV}$} and \textsc{very shallow-$\rho_{\rm UV}$} models, respectively.  The vertical dotted line denotes $z = 17$.  

Models with higher UV densities, fainter $M_{\rm UV}^{\rm cut}$, and higher $f_{\rm esc}^{\rm X}$ produce larger IGM temperatures.  All these scenarios begin pre-heating the IGM at $z > 17$, and some of them have already reached gas temperatures above $T_{\rm CMB}$ by this redshift.  This indicates that X-ray pre-heating by AGN can easily suppress the 21 cm absorption signal, working against the effect of the AGN radio emission studied in this work.  Our models with $f_{\rm esc}^{\rm X} = 1$ produce IGM temperatures much higher than expected for pre-heating by X-ray binaries~\citep{Fialkov2014c,Eide2018}, with $T_{\rm K}$ reaching $\gtrsim 10^4$K by\footnote{Models with this much pre-heating may even violate measurements of the IGM temperature at $z < 6$ from the Ly$\alpha$ forest, which find $T \sim 1-1.5 \times 10^4$ K after reionization~\citep[e.g.][]{Boera2019,Gaikwad2020}.  } $z = 10$  These findings indicate that strong 21 cm absorption signals may be difficult to produce in our model unless the production and/or escape of X-ray photons from $z > 15$ AGN is suppressed.  
 
\section{Results}
\label{sec:results}

\subsection{Minimum requirements for the EDGES depth}
\label{subsec:minimum}

\begin{figure*}
    \centering
    \includegraphics[scale=0.385]{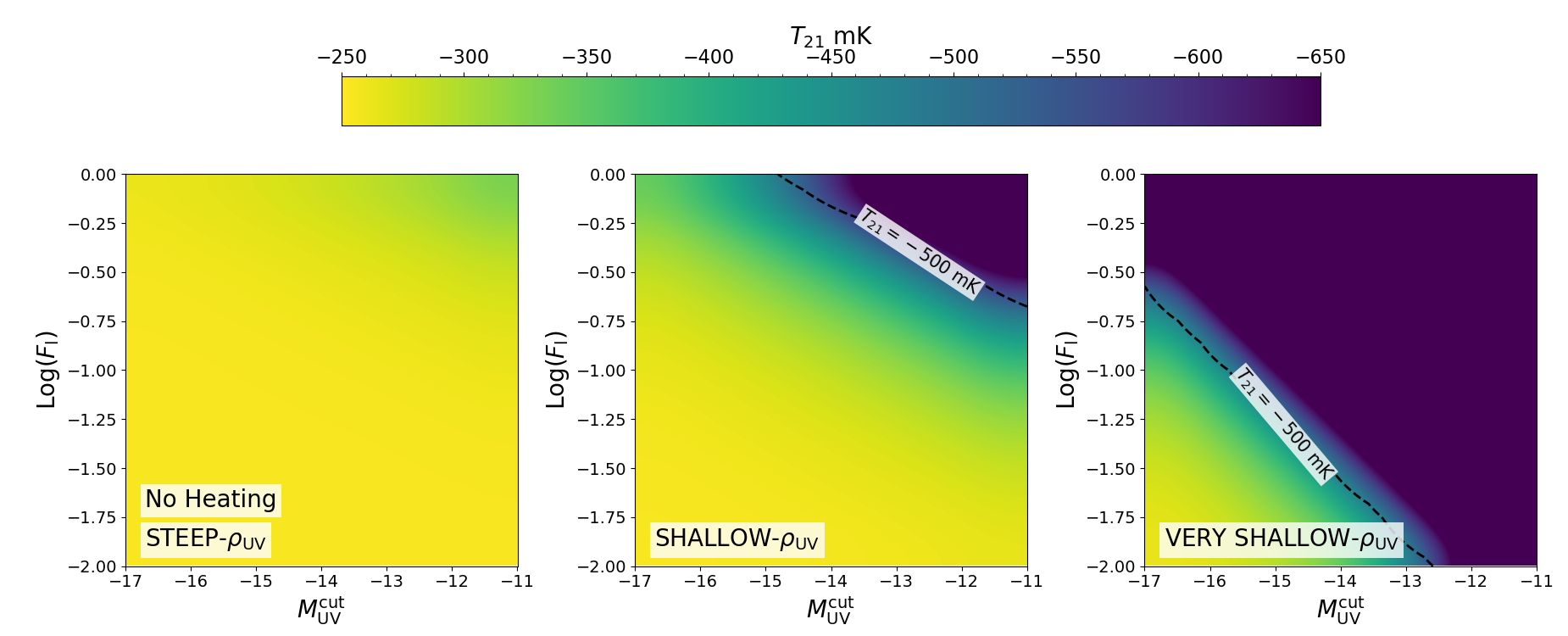} 
    \caption{Minimum properties of high-redshift AGN required to reach $T_{21} = -500$ mK by $z = 17$.  We show $T_{21}$ at $z = 17$ vs. $\log(F_{\rm l})$ (vertical axis) and $M_{\rm UV}^{\rm cut}$ (horizontal axis), with the dashed lines denoting $T_{21} = -500$ mK.  These calculations assume complete Ly$\alpha$ coupling and no X-ray heating, the most optimistic assumptions for the signal depth.  No part of our parameter space in the \textsc{steep-$\rho_{\rm UV}$} model can reach $-500$ mK by $z = 17$.  The \textsc{shallow-$\rho_{\rm UV}$} model (middle panel) only does so in the top right corner of parameter space, where nearly all objects down to very faint magnitudes ($M_{\rm UV} \gtrsim -14$) are radio-loud AGN.  Only the \textsc{very shallow-$\rho_{\rm UV}$} model (right-most panel) can achieve the required depth with $M_{\rm UV}^{\rm cut} \leq -17$ {\it or} a radio-loud fraction $< 10\%$.  }
    \label{fig:minimum}
\end{figure*}

We begin by considering the minimum requirements to produce a 21 cm signal as deep as that measured by EDGES, $T_{21} = -500$ mK, by $z = 17$.  The depth of the absorption signal is maximized if (1) X-ray preheating has not started, such that $T_{\rm K}$ is as small as possible, and (2) Ly$\alpha$ coupling is already complete, so that $T_{\rm S} = T_{\rm K}$ (see Equation ~\ref{eq:T21}-\ref{eq:Ts}).  In this section, we assume these two conditions, and consider what part of our parameter space results in $T_{21} \leq -500$ mK by $z = 17$.  

The three most important parameters that affect radio emission in our model are $f_{\rm bh}$, $f_{\rm l}$, and $M_{\rm UV}^{\rm cut}$ (see Equation ~\ref{eq:epsilon_radio}).  The degenerate combination $f_{\rm bh} f_{\rm l}$, which we denote $F_{\rm l}$, is the fraction of {\it all} objects with $M_{\rm UV} < M_{\rm UV}^{\rm cut}$ that are radio-loud AGN.  The Eddington ratio $\eta_{\rm edd}$ also enters Equation ~\ref{eq:epsilon_radio} through the conversion from UV luminosity to black hole mass in Equation ~\ref{eq:Lbh} and the subsequent conversion to radio luminosity in Equation ~\ref{eq:L5}-\ref{eq:L_2_10}.  We find that the dependence of the excess radio background on $\eta_{\rm edd}$ is relatively weak - namely, changing $\eta_{\rm edd}$ by two orders of magnitude only changes $T_{21}$ by a factor of $\sim 2$.  This behavior occurs because $\eta_{\rm edd}$ enters through both Equations ~\ref{eq:Lbh} and~\ref{eq:L_2_10}, and almost cancels out in the conversion from $L_{\rm UV}$ to $L_{
\rm 5GHz}$.  Following~\citet{Hegde2024}, we (optimistically) set $\eta_{\rm edd} = 1$, but note that our results are not strongly sensitive to this choice.

In Figure~\ref{fig:minimum}, we show $T_{\rm 21}(z = 17)$ vs. $\log(F_{\rm l})$ (vertical axis) and $M_{\rm UV}^{\rm cut}$ (horizontal axis) for each of our three UVLF models assuming an adiabatically cooled IGM and complete Ly$\alpha$ coupling.  In each panel, the dashed line denotes the iso-contour corresponding to $T_{21} = -500$ mK.  Parts of parameter space above this line - with higher $F_{\rm l}$ and/or fainter $M_{\rm UV}^{\rm cut}$ (dark blue regions) - have the potential to reach the absorption depth required by EDGES.  Conversely, the green/yellow regions below this line cannot produce EDGES even under the most optimistic circumstances.  
We see in the left-most panel that no part of parameter space in the \textsc{steep-$\rho_{\rm UV}$} model can reach $-500$ mK by $z = 17$.  Even if {\it every} UV-bright object were a radio-loud AGN at $z \geq 17$ down to $M_{\rm UV} = -11$ in this model, their combined radio output would be insufficient to backlight EDGES.  This model is most consistent with pre-JWST expectations for the UVLF~\citep{Oesch2018} and models that assume constant star formation efficiency in dark matter halos well above $z = 10$~\citep{Harikane2023,Donnan2025}.

The \textsc{shallow-$\rho_{\rm UV}$} and \textsc{very shallow-$\rho_{\rm UV}$} models both produce radio backgrounds large enough to backlight the EDGES depth in some part of our parameter space.  The former requires that either $\approx 20\%$ of all objects brighter than $M_{\rm UV} = -11$ and/or all objects with $M_{\rm UV} < -14$ are radio-loud AGN.  This is in contrast to observations at lower redshifts, which find that the transition in the UVLF between AGN and stellar-dominated objects should occur at much brighter $M_{\rm UV}$~\citep[e.g.][]{Finkelstein2022}.  A radio-loud fraction near unity is also unexpected, since radio observations of quasars at $z \lesssim 6$ find radio-loud fractions of order $\sim 10\%$ or less~\citep{Banados2015,Liu2021,Keller2024}. 
In our \textsc{very shallow-$\rho_{\rm UV}$} model, $F_{\rm l} < 10\%$ is still able to achieve the EDGES depth if $M_{\rm UV}^{\rm cut} < -16$, and $M_{\rm UV}^{\rm cut} < -17$ is allowed if the radio loud fraction is at least $25\%$.   However, this model requires minimal evolution (an order of magnitude or less) in the UVLF deep into the CD, which would be challenging to reconcile with even the most optimistic galaxy/AGN formation models and observational limits~\citep{Yung2025,Castellano2025,PerezGonzalez2025}.  We note that including the effects of Ly$\alpha$ heating would make this result even more pessimistic.    

These results show that even under the most optimistic conditions, explaining the depth of the EDGES signal with radio-loud AGN is challenging.  Such a scenario requires at least two of three conditions to hold: (1) that the UV luminosity density evolves more gradually than expected at $z > 15$, (2) that $\gtrsim 20\%$ of AGN are radio-loud, in contrast to lower redshift observations, and/or (3) that the majority of objects as faint as $M_{\rm UV} = -11$ have their UV light dominated by AGN.  In the rest of this work, we will explore the additional conditions required to realize the depth and shape EDGES signal in such a model under less optimistic physical assumptions.  

\subsection{Additional conditions to achieve the EDGES depth}
\label{subsec:depth}

 \begin{figure}
    \centering
    \includegraphics[scale=0.45]{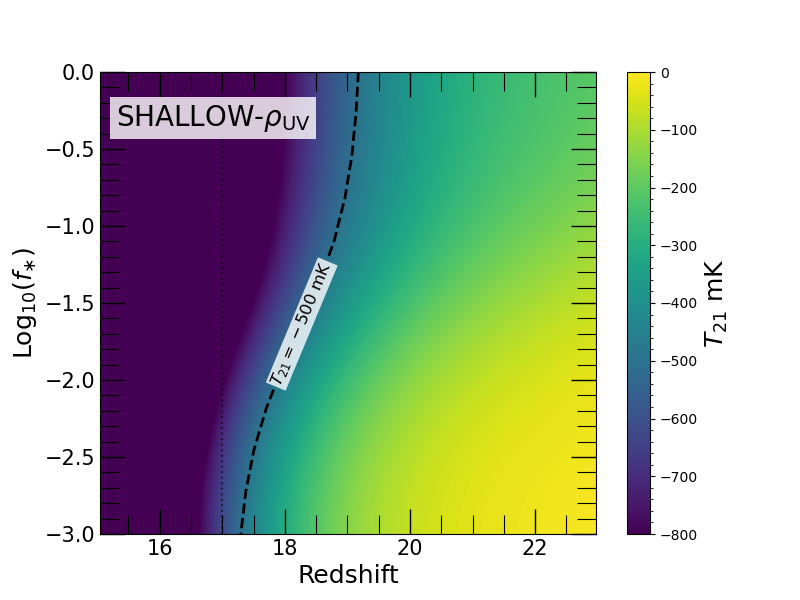} 
    \caption{Effect of Ly$\alpha$ coupling on the evolution of $T_{21}$ in our \textsc{shallow-$\rho_{\rm UV}$} model. 
 We show $T_{21}$ as a function of $f_{\ast}$ (vertical) and redshift (horizontal) for our most optimistic scenario ($M_{\rm UV}^{\rm cut} = -11$ and $f_{\rm l} = f_{\rm bh} = 1$).  The dashed line denotes $T_{21} = -500$ mK, as in Figure~\ref{fig:minimum}.  The redshift at which $T_{21} = -500$ mK is reached shifts from $z = 19.3$ for $f_{\ast} = 1$ to $z = 17.5$ when $f_{\ast} = 0.001$.  Even our \textsc{shallow-$\rho_{\rm UV}$} model cannot reach the required EDGES depth without the help of pop III stars to initiate an early onset to Ly$\alpha$ coupling (see text for details).  }
    \label{fig:fstar}
\end{figure}

\begin{figure*}
    \includegraphics[scale=0.60]{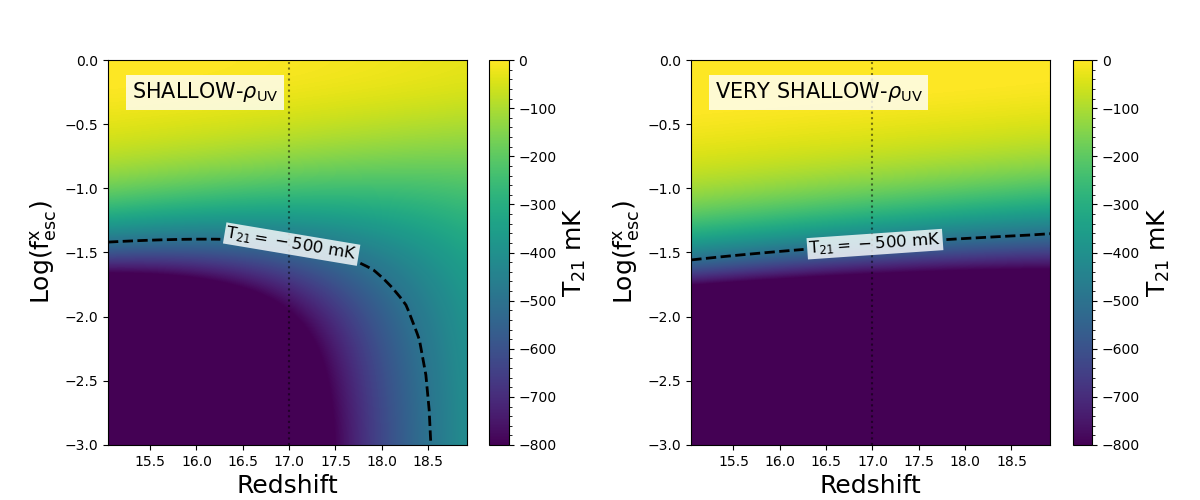} 
    \caption{Conditions on $f_{\rm esc}^{\rm X}$ required to achieve the EDGES absorption depth.  The left and right panels show the \textsc{shallow-$\rho_{\rm UV}$} and \textsc{very shallow-$\rho_{\rm UV}$} models, respectively, and the format is the same as that of Figure~\ref{fig:fstar}.  We assume $F_{\rm L} = 1$ and $M_{\rm UV}^{\rm cut} = -11$, as in Figure~\ref{fig:fstar}.  We find that \revision{$\log(f_{\rm esc}^{\rm X}) \lesssim -1.5$} is required to reach $T_{21} = -500$ mK by $z \sim 17$ on the left and right.  In the \textsc{shallow-$\rho_{\rm UV}$} model, the dashed line drops towards $f_{\rm esc}^{\rm X} = 0$ approaching $z = 19$, while in the \textsc{very shallow-$\rho_{\rm UV}$} case it remains roughly constant with redshift.  }
    \label{fig:fesc_contour}
\end{figure*}

In this section, we include realistic treatments of Ly$\alpha$ coupling and X-ray heating, and study the conditions required to explain the EDGES depth in our \textsc{shallow-$\rho_{\rm UV}$} and \textsc{very shallow-$\rho_{\rm UV}$} models.  In Figure~\ref{fig:fstar}, we illustrate how Ly$\alpha$ coupling affects the depth of the signal by varying the pop III star formation efficiency, $f_{\ast}$ (Equation~\ref{eq:rhoSFRIII}).  The color-scale shows $T_{21}$ as a function of $f_{\ast}$ (vertical axis) and redshift (horizontal axis), and the dashed line denotes $T_{21} = -500$ mK, as in Figure~\ref{fig:minimum}. The vertical dashed line denotes $z = 17$. 
 For this model, we assume $f_{\rm l} = f_{\rm bh} = 1$ and $M_{\rm UV}^{\rm cut} = -11$, our most optimistic combination of parameters.  In the \textsc{shallow-$\rho_{\rm UV}$} model, lower values of $f_{\ast}$ delay the onset of Ly$\alpha$ coupling, when $x_{\alpha} \sim 1$, to lower redshifts (see top panel of Figure~\ref{fig:xalpha}).  This, in turn, delays the redshift at which $T_{21}$ reaches the EDGES depth from $z = 19.3$ (for $f_{\ast} = 1$) to $z = 17.5$ ($f_{\ast} = 0.001$).  At very low $f_{\ast}$, the UV photons produced by AGN and pop II stars become the dominant source of Ly$\alpha$ coupling.  In this limit, the \textsc{shallow-$\rho_{\rm UV}$} model barely reaches the EDGES depth by $z = 17$ in the most optimistic scenario, since UV photons from AGN and pop II stars complete Ly$\alpha$ coupling later than Pop III stars.  This is especially problematic given that the flat-bottomed EDGES trough extends to $z \sim 19.5$, requiring an earlier completion of Ly$\alpha$ coupling.  We note that a smaller value of $M_{\rm crit}$ (see discussion in \S\ref{subsubsec:popIII}) would help relax the requirements on $f_{\ast}$.

By contrast, the bottom panel of Figure~\ref{fig:xalpha} shows that in the \textsc{very shallow-$\rho_{\rm UV}$} case, $x_{\alpha}$ is nearly independent of $f_{\ast}$, since the bulk of UV photons are produced by AGN and pop II stars.  Indeed, in that model the UV production is so high that $x_{\alpha} = 1$ is achieved very quickly after $z = 30$, and would occur much earlier if we did not impose the condition that no UV photons are produced at $z > 30$.  In this model, AGN drive both the coupling of $T_{\rm S}$ to the gas temperature and the buildup of the radio background that backlights the signal, such that pop III stars play a minimal role.  We thus find that relatively early Ly$\alpha$ coupling by pop III stars is necessary for models like our \textsc{shallow-$\rho_{\rm UV}$} case to achieve the EDGES depth.  Scenarios with higher $\rho_{\rm UV}$ from AGN and pop II stars at $z > 15$ can produce a stronger radio background and achieve Ly$\alpha$ coupling by $z = 17$ without the help of pop III stars.  

As outlined in \S\ref{subsec:xray_heating}, AGN are expected to be copious X-ray producers as well as sources of radio photons.  X-ray photons pre-heat the IGM, raising $T_{\rm S}$ and decreasing the amplitude of the absorption signal.  Avoiding this requires that either the X-ray escape fraction ($f_{\rm esc}^{\rm X}$ in Equation~\ref{eq:NdotX}) is small, and/or that high-redshift AGN produce fewer high-energy photons than their low-redshift counterparts.  The former could be due to e.g. obscuration in the early growth phases of AGN~\citep{Yang2023}, and the latter may be expected if these super-Eddington accretion  is common at these redshifts~\citep{Madau2024}.  Here, we use $f_{\rm esc}^{\rm X}$ as a proxy for both scenarios.  

In Figure~\ref{fig:fesc_contour}, we show the conditions on $f_{\rm esc}^{\rm X}$ required to achieve the EDGES absorption depth, in the same format as Figure~\ref{fig:fstar}.  We assume $f_{\ast} = 0.05$, as in Figure~\ref{fig:xalpha}, continue using $F_{\rm l} = 1$ and $M_{\rm UV}^{\rm cut} = -11$.  The left and right panels show our \textsc{shallow-$\rho_{\rm UV}$} and \textsc{very shallow-$\rho_{\rm UV}$} models.  We see that they require $\log(f_{\rm esc}^{\rm X}) \approx -1.4$ to achieve the EDGES depth at $z = 17$.  This is roughly an order of magnitude lower than {\it Lyman Continuum} (LyC) escape fractions of $f_{\rm esc}^{\rm LyC} \sim 0.3$ measured for faint AGN at $z \lesssim 4$ measured by~\citet{Smith2020}.  This discrepancy is exacerbated by the fact that the absorption cross-section of HI is several orders of magnitude smaller for hard X-rays than for LyC photons, suggesting that X-rays should escape AGN more easily than LyC photons (such that $f_{\rm esc}^{\rm X} \geq f_{\rm esc}^{\rm LyC}$).  Indeed, for the shape of the X-ray spectrum assumed in this work, this corresponds to HI column densities\footnote{The effective column density needed to produce an escape fraction of $f_{\rm esc}^{\rm X}$ is given implicitly by $f_{\rm esc}^{\rm X} = \left[\int_{\nu_{0.5}}^{\nu_{10}} d\nu\dot{N}_{\rm X}(\nu,z) \exp[-N_{\rm HI}\sigma_{\rm eff}(\nu)]\right]/\left[\int_{\nu_{0.5}}^{\nu_{10}} d\nu\dot{N}_{\rm X}(\nu,z)\right]$, where $\sigma_{\rm eff} \equiv [\sigma_{\rm HI}(\nu)N_{\rm HI} + \sigma_{\rm He}(\nu)\chi N_{\rm HI}]/N_{\rm HI}$, with $\chi = 0.082$ to correct for helium. All other quantities are defined as they are in Equations~\ref{eq:NdotX}-\ref{eq:HXray}.  This equation assumes a uniform shell of HI gas around the AGN, while an alternative interpretation would take $f_{\rm esc}^{\rm X}$ to one minus the covering fraction of high-column HI in a ``porous shell'' geometry.  } of $\log(N_{\rm HI}/{\rm cm}^{-2}) \approx 23.82$, corresponding to highly obscured AGN~\citep{Gilli2022}.  At $z \approx 19$, the maximum $f_{\rm esc}^{\rm X}$ approaches $0$ for the \textsc{shallow-$\rho_{\rm UV}$} model, since that model cannot achieve the EDGES depth by that redshift even in the most optimistic case.  In the \textsc{very shallow-$\rho_{\rm UV}$}, it remains nearly constant with redshift.  

It is interesting that the maximum value allowed for $f_{\rm esc}^{\rm X}$ to produce the EDGES depth is similar in these two models, despite the fact that the \textsc{very shallow-$\rho_{\rm UV}$} case produces a much stronger radio background (see Figure~\ref{fig:minimum}).  This insensitivity arises from the fact that the same AGN producing the radio background are also the X-ray producers, such that the ratio of radio to X-ray output does not depend too strongly on the assumed UVLF.  Indeed, this coupling is a direct consequence from our use of the quasar fundamental plane to estimate the radio output of AGN directly from their X-ray luminosities (Equation ~\ref{eq:L5}).  Thus, a stronger radio background will always be accompanied by more X-ray heating, and the only way to significantly change the relationship between these is via $f_{\rm esc}^{\rm X}$.  Thus, we conclude that an AGN-driven backlight for the EDGES signal demands very low X-ray escape fractions and/or much weaker X-ray production than analogous AGN at lower redshifts, by at least an order of magnitude.  

\begin{figure}[hbt!]
    \centering 
    \includegraphics[scale=0.42]{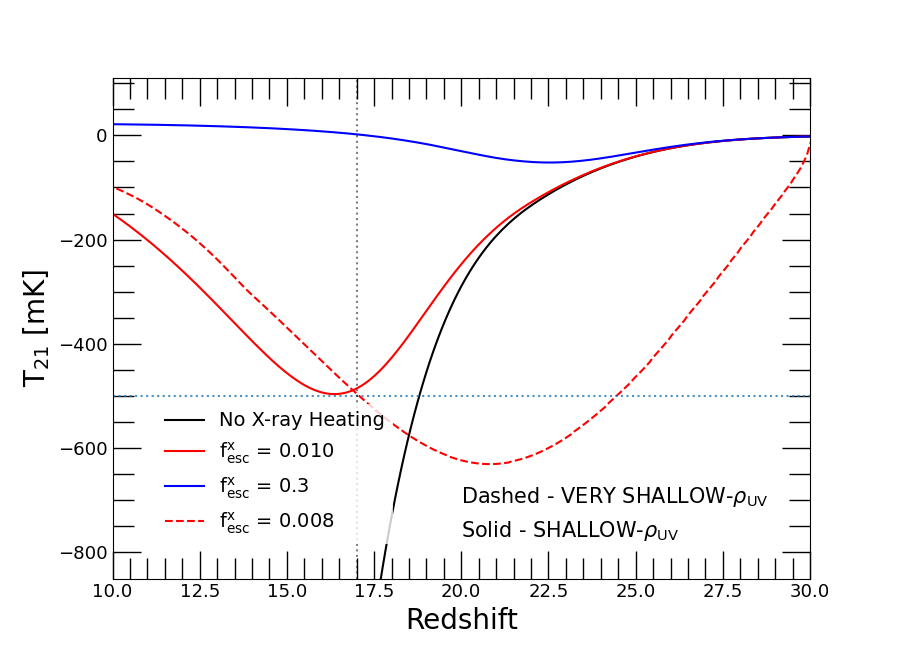}
    \caption{Examples of scenarios that produce the $T_{21} = -500$ mK absorption depth seen by EDGES in the \textsc{shallow-$\rho_{\rm UV}$} (red solid) and \textsc{very shallow-$\rho_{\rm UV}$} (red dashed) models.  Both models include Ly$\alpha$ coupling (with $f_{\ast} = 0.05$).  The \textsc{shallow-$\rho_{\rm UV}$} and \textsc{very shallow-$\rho_{\rm UV}$} assume $f_{\rm esc}^{\rm X} = 0.04$ and $0.0375$, respectively.  The black solid curve shows what the \textsc{shallow-$\rho_{\rm UV}$} case would look like without any X-ray heating, and the blue solid curve shows the same for a much higher $f_{\rm esc}^{\rm X} = 0.3$.  The \textsc{shallow-$\rho_{\rm UV}$} model reaches the target depth slightly too late at $z \approx 16$, and the \textsc{very shallow-$\rho_{\rm UV}$} case, though it goes through $-500$ mK at $z = 17$, reaches it too early at $z \approx 21$.  See text for discussion.  }
    \label{fig:T21_heating_plot}
\end{figure}

In Figure~\ref{fig:T21_heating_plot}, we show examples of scenarios that can reach $T_{21} = -500$ mK in the $15 < z < 20$ range.  The solid and dashed red curves show $T_{21}$ vs. $z$ for our \textsc{shallow-$\rho_{\rm UV}$} and \textsc{very shallow-$\rho_{\rm UV}$} models, respectively, for scenarios that include the effects of Ly$\alpha$ coupling (with $f_{\ast} = 0.05$) and X-ray heating.  For the former, we assume $F_{\rm l} = 1$ and $M_{\rm UV}^{\rm cut} = -11$, and the latter we use less extreme values of $F_{\rm l} = 0.85$ and $M_{\rm UV}^{\rm cut} = -15$.  Respectively, the two models assume $f_{\rm X}^{\rm esc} = 0.04$ and $0.0375$, close to the dashed lines in Figure~\ref{fig:fesc_contour}.  The black solid curve shows the \textsc{shallow-$\rho_{\rm UV}$} case in the absence of any X-ray heating, and the blue solid curve shows the same for a much higher $f_{\rm esc}^{\rm X} = 0.3$, consistent with measurements of $f_{\rm esc}^{\rm LyC}$ by~\citet{Smith2020} at much lower redshift.  The vertical dotted line indicates $z = 17$ and the horizontal one denotes $-500$ mK.  

We see that the \textsc{shallow-$\rho_{\rm UV}$} model reaches the depth of the EDGES signal slightly too late, at $z \approx 16$ rather than central $z = 17$.  This is because X-ray heating, which is needed to produce the up-turn in the signal at lower redshifts, starts at $z \approx 20$ and reduces the depth of the absorption feature slightly at $z = 17$.  This up-turn is fairly gradual in our model, with $T_{21} = -150$ mK by $z = 10$.  This is due to the ongoing production of radio photons at $z < 15$, which makes it more difficult for the signal to saturate (that is, to reach $T_{\rm S} >> T_{\rm radio}$).  The blue solid curve shows that much higher X-ray escape washes out the CD signal almost completely.  The \textsc{very shallow-$\rho_{\rm UV}$} model, by contrast, goes through $-500$ mK at $z = 17$, but reaches the EDGES depth early at $z \approx 21$, thanks to the rapid early buildup of the radio background and onset of Ly$\alpha$ coupling.  In this case, the signal also takes a long time to saturate, reaching $T_{21} = -100$ mK by $z = 10$.  

Recently, an additional heating mechanism driven by the soft radio photons themselves - soft-photon heating - was pointed out by~\citet{Acharya2023,Cyr2024}.  They found that for synchrotron-like spectra ($\alpha_{\rm radio} \sim 0.6$), this heating source could increase the gas temperature enough to cancel out the effect of the enhanced background on the 21 cm signal, resulting in the same or an even lower signal than expected without radio emission.  Since we neglect this effect here, the true prospects to achieve the EDGES depth are likely more pessimistic than we find here, to a degree that depends on the value of $\alpha_{\rm radio}$.  If $\alpha_{\rm radio} = 0.5$, our fiducial assumption, soft photon heating may render the EDGES depth impossible to achieve.  However, if $\alpha_{\rm radio} \approx 0$, as for typical core-dominated AGN~\citep{deGasperin2018}, soft photon heating would likely not dominate the total heating rate.  As mentioned in \S\ref{subsec:radio_AGN}, since our radio fluxes are based on measurements close to the target frequency of $1.4$ GHz, our results are reasonably insensitive to the true value of $\alpha_{\rm radio}$.  As such, an EDGES signal back-lit by AGN radio emission is likely still possible, even accounting for soft photon heating, but would carry the additional requirement of flat radio spectra.  

Our findings suggest that achieving the EDGES depth at $z \sim 17$ with a radio background produced by AGN is challenging even under optimistic assumptions about the radio emission properties of the AGN population.  Scenarios with low UV production at $z > 20$ require a significant early contribution to Ly$\alpha$ coupling from pop III stars to lower $T_{\rm S}$ enough to achieve the required depth.  In all scenarios, the X-ray escape fraction and/or intrinsic X-ray production of AGN must be at least $\approx 1.5$ orders of magnitude below low-redshift expectations to avoid washing out the signal via pre-heating.  These findings further narrow the parameter space within which AGN could provide the necessary radio backlight to explain EDGES.  

\subsection{Conditions to reproduce the shape of EDGES}
\label{subsec:shape}

In this section, we consider the conditions required for AGN to explain the unexpected shape of the EDGES signal, and not just its depth.  The flattened bottom and steep redshift evolution on both sides of the EDGES signal are unexpected in canonical models of the 21 cm CD signal, which prefer less abrupt redshift evolution~\citep{Cang2024}.  Recently,~\citet{Mittal2022} showed that reproducing these features of EDGES in their model required sharp redshift evolution in the properties of the galaxy population that would be difficult to explain physically.  We re-visit this problem in the context of our AGN-driven model here.  

One key feature of EDGES is its sharp evolution from near-zero signal at $z \approx 21.5$ to its maximum depth of $T_{21} = -500$ mK at $z \approx 19$, over a redshift interval of just $\Delta z \approx 2.5$.  This is hard to reproduce in our models, which generally predict longer redshift intervals ($\Delta z \approx 5-10$) between the onset of the signal and maximum depth (see Figure~\ref{fig:T21_heating_plot}).  There are two plausible mechanisms that could drive such a fast transition in our model: (1) either Ly$\alpha$ coupling happens more rapidly than expected, and/or (2) the excess radio background builds up rapidly $z < 21.5$.  The problem with (2) is that even absent any excess radio background before $z = 21.5$, the CMB alone would be sufficient to backlight absorption at these redshifts if Ly$\alpha$ coupling was already underway.  
We have explored both types of scenarios and find that the high-redshift ``edge'' of the signal is impossible to fit without an abrupt shutoff of {\it all} UV sources (AGN, Pop II, and Pop III stars) at $z > 21.5$.  Such a scenario might be achievable, for example, if the most massive halos at this epoch dominate the UVLF~\citep{Kaurov2019}. Note that in our \textsc{very shallow-$\rho_{\rm UV}$}, Ly$\alpha$ coupling is driven by AGN and Pop II stars rather than Pop III (see Figure~\ref{fig:xalpha}), such that the main effect of this cutoff is to set $f_{\rm bh} = 0$ at $z > 21.5$.  

We turn next to the rapid saturation of the EDGES signal between $z \approx 16$ and $14$.  Assuming Ly$\alpha$ coupling is complete by this redshift, we could try to explain this feature by adding redshift evolution to any of our parameters that control the growth of the radio background and/or heating by X-rays.  There are two plausible ways to do this: (1) the radio background stops growing at $z \approx 16$, allowing $T_{\rm S}$ to catch up quickly with $T_{\rm radio}$~\citep{Mebane2019}, and/or (2) X-ray heating increases rapidly at $z < 16$, increasing $T_{\rm S}$ such that the same result is achieved.  Evidence for (1) may already have some observational support.  Recently,~\citet{Dsilva2023} performed SED fitting on a sample of high-redshift galaxies from the CEERS~\citep{Finkelstein2023} and JWST-GLASS~\citep{Treu2022} surveys, including fits that allowed some of the observed UV light to originate from AGN.  
They found that at $z \approx 5$ and $7.5$, they obtained similar results with and without an AGN component in their fits.  However, at $z \approx 10$, their recovered star formation rate was lower when including the AGN component by up to a factor of $\approx 2$, suggesting that AGN could be responsible for up to half the observed UV light at that redshift.  
This is consistent with the results of~\citet{Hegde2024}, who found that up to half of the light from $z > 10$ objects could originate from AGN sources without violating constraints on galaxy morphology.  

One interpretation of the~\citet{Dsilva2023} results is that they indicate evolution in the brightness cutoff between stellar-dominated and AGN-dominated sources (that is, $M_{\rm UV}^{\rm cut}$).  Assuming a sharp cutoff between the two populations, the fraction of UV light produced by AGN can be expressed as

\begin{equation}
    \label{eq:MUVcut_DSilva}
    x_{\rm AGN} = \frac{\rho_{\rm UV}^{\rm AGN}}{\rho_{
    \rm UV}^{\rm total}} = \frac{\int_{-\infty}^{\rm M_{\rm UV}^{\rm cut}}dM_{\rm UV} \frac{dn}{dM_{\rm UV} }L_{\rm UV}}{\int_{-\infty}^{-17}dM_{\rm UV} \frac{dn}{dM_{\rm UV} }L_{\rm UV}}
\end{equation}


\begin{figure}[hbt!]
    \centering 
    \includegraphics[scale=0.46]{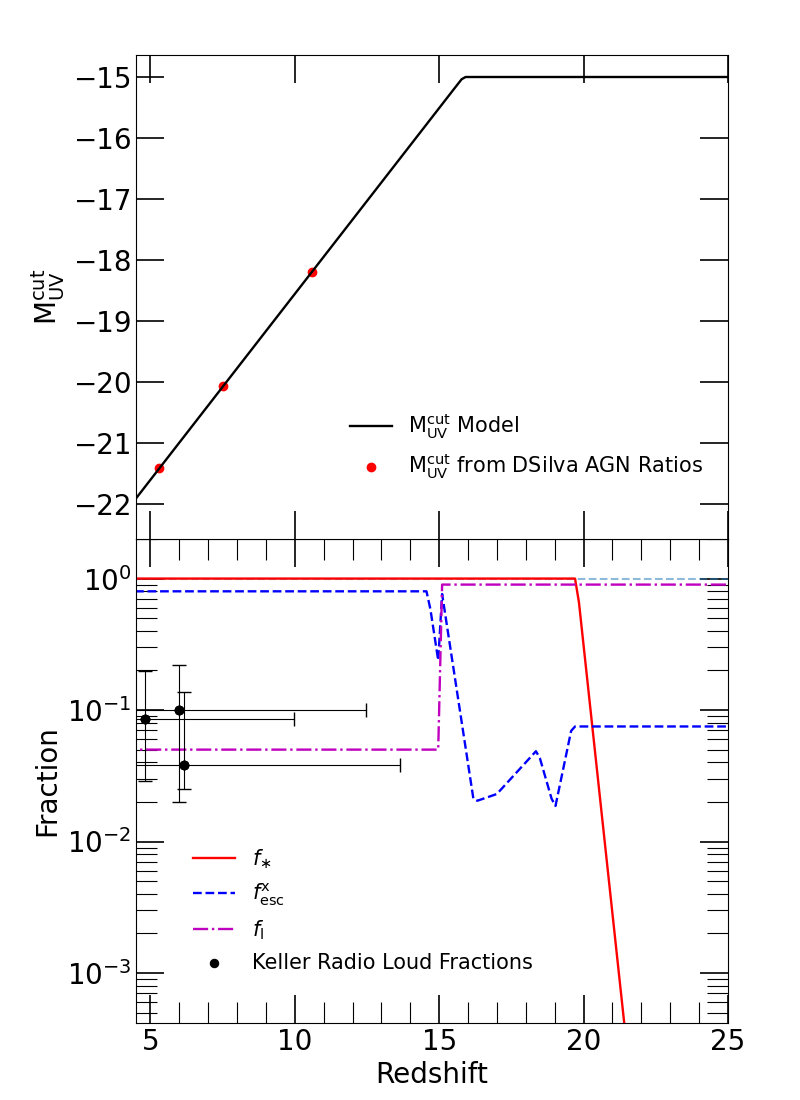}
    \caption{Evolution of physical quantities in the scenarios we explore with redshift-dependent parameters.  {\bf Top:} our model for evolution in $M_{\rm UV}^{\rm cut}$, motivated by estimates of the fraction of UV light produced by AGN at $5 < z < 11$ from~\citet{Dsilva2023}.  The red points are based on measurements from that work (see text), and the black line shows our extrapolation to higher redshifts (capped at $M_{\rm UV}^{\rm cut} = -15$).  {\bf Bottom:} evolution of $f_\ast$ (red), $f_l$ (magenta) and $f_{\rm esc}^{\rm X}$ (blue) assumed in this section.  The black points show recent measurements of the radio-loud fraction of bright quasars at $5 < z < 6$, to which our evolving $f_l$ model is anchored.  See text for details.  }
    \label{fig:evolution}
\end{figure}

where the upper limit of $M_{\rm UV} = -17$ in the denominator reflects the cutoff used to define $\rho_{\rm UV}$ in observations (e.g.~\citet{Adams2023}).  If we take the ratio of the red and blue points in the Figure 3 of~\citet{Dsilva2023} to be the fraction of light produced by stars ($1 - x_{\rm AGN}$), we can estimate $M_{\rm UV}^{\rm cut}$ from their data.  We perform this exercise and plot the results in the bottom panel of Figure~\ref{fig:evolution}.  The three red points denote the values of $M_{\rm UV}^{\rm cut}$ inferred in this way from~\citet{Dsilva2023} at $z = 5.5$, $7.5$, and $10.5$, which increase linearly with redshift.  The black solid line shows a linear extrapolation of these points to higher redshifts, which we cap at $z = 16$ ($M_{\rm UV}^{\rm cut} = -15$).  An important caveat is that these results are subject to limited number statistics, especially in the $z = 10.5$ bin, and should thus be treated with some caution.  

We can see that producing enough radio photons to explain EDGES would require significant redshift evolution in $f_l$ between $z = 6$ and $15$.  Measurements of $f_l$ at $5 < z < 6$ find values in the $5-10\%$ range for bright quasars~\citep{Banados2015,Liu2021,Keller2024}, far below the $\gtrsim 25\%$ values required in most of our parameter space to explain EDGES.  A rapid transition between $f_l = 1$ and $0.05$ around $z = 15$ would respect these low-$z$ boundary conditions whilst maximally accelerating the saturation of the signal at $z < 15$.  We show an extreme (step-function) case of such a model as the magenta curve in the bottom panel of Figure~\ref{fig:evolution}.  The black points denote the measured radio-loud fractions at $5 < z < 6$ compiled by~\citet{Keller2024}.  The other possibility is that evolution in $f_{\rm esc}^{\rm X}$ drives saturation at $z \sim 15$.  We showed in Figure~\ref{fig:fesc_contour} that $f_{\rm esc}^{\rm X} \sim 5\%$ ($\log(N_{\rm HI}/{\rm cm}^{-2}) \sim 24.7$) is required to produce the EDGES signal depth.  However, LyC escape fractions are observed to be $\sim 30\%$ at $z \sim 3$~\citep{Smith2020}.  Allowing $f_{\rm esc}^{\rm X}$ to evolve rapidly to $30\%$ at $z = 15$ could also explain the rapid saturation.  We show such a model as the blue curve in Figure~\ref{fig:evolution} (see below for details).

\begin{figure}[hbt!]
    \centering
    \includegraphics[scale=0.44]{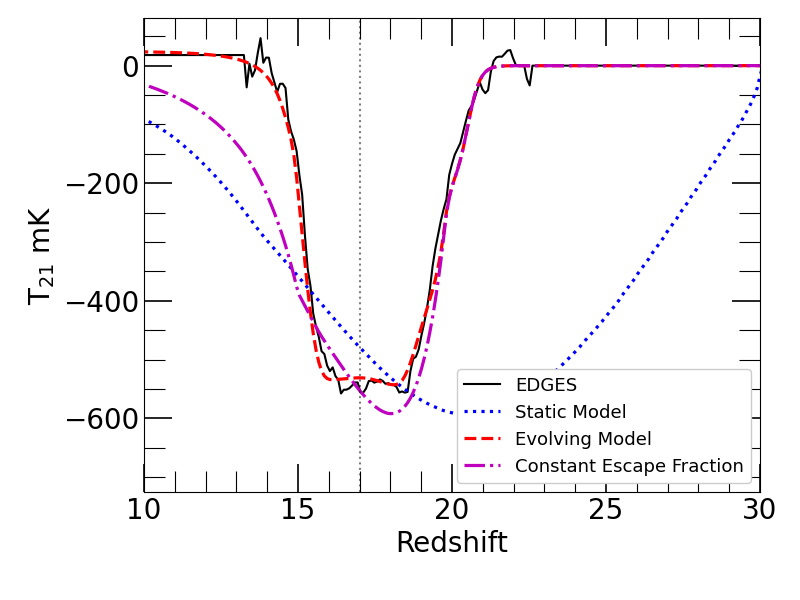}
    \caption{Models of $T_{21}$ with redshift-dependent parameters, compared against the EDGES signal (black solid).  The blue dotted curve is our \textsc{very shallow-$\rho_{\rm UV}$} model with no redshift-dependent parameters from Figure~\ref{fig:T21_heating_plot}.  The magenta dot-dashed curve assumes a sharp cutoff in UV emission at $z > 21.5$, and the observationally-motivated models for evolving $M_{\rm UV}^{\rm cut}$ and $f_l$ shown in Figure~\ref{fig:evolution}, but retains a constant $f_{\rm esc}^{\rm X}$.  This case comes much closer to the shape of the EDGES signal, but still misses the rapid saturation at $z \approx 15$ and the flat-bottomed shape of the signal.  The red curve uses the evolving $f_{\rm esc}^{\rm X}$ model in Figure~\ref{fig:evolution}.  By fine-tuning the shape of $f_{\rm esc}^{\rm X}(z)$, we were able to obtain a good fit to the shape of EDGES.  }
    \label{fig:best_fit}
\end{figure}

In Figure~\ref{fig:best_fit}, we show how allowing parameters to evolve affects the signal.  The blue dotted curve is the same as the red dashed curve in Figure~\ref{fig:T21_heating_plot}, which assumes constant parameters.  The dot-dashed curve includes a sharp cutoff in UV production at $z > 21.5$, to achieve rapid onset of Ly$\alpha$ coupling, and our observationally-motivated models for the redshift evolution of $M_{\rm UV}^{\rm cut}$ and $f_l$ shown in Figure~\ref{fig:evolution}, but a constant $f_{\rm esc}^{\rm X}$ that is the same as the blue dotted curve.  The red dashed curve adds the evolving $f_{\rm esc}^{\rm X}$ model given by the blue curve in Figure~\ref{fig:evolution}, which we have fine-tuned to fit the observed EDGES signal (black solid).  
We see that although evolving $M_{\rm UV}^{\rm cut}$ and $f_l$ saturates the signal faster than the ``static'' case, the absorption profile remains broader than EDGES.  This is because a large radio background remains in place even after the sources of radio emission shut off, preventing the signal from vanishing quickly.  By contrast, a rapid increase in $f_{\rm esc}^{\rm X}$ drives the IGM temperature well above $T_{\rm radio}$ by $z = 14$.  Several previous works have noted that similarly extreme evolution in astrophysical properties of early sources are required to reproduce these features~\citep{Fialkov2019,Mittal2022}.  

We also find that fine-tuning the redshift evolution of $f_{\rm esc}^{\rm X}$ at $15 < z < 20$ allows us to also reproduce the flat-bottomed shape of the signal.  Note that the shape of the dot-dashed magenta curve in Figure~\ref{fig:best_fit} has a decreasing signal amplitude between its minimum at $z = 18$ and $z = 15$, which does not match EDGES.  The local maximum at $z = 18$ in our evolving $f_{\rm esc}^{\rm X}$ model (blue dashed curve in Figure~\ref{fig:evolution}) serves to flatten out this redshift evolution, essentially ``canceling out'' a well-defined minimum in the signal in this redshift range. Thus, we see that the jagged shape of the $f_{\rm esc}^{\rm X}$ evolution in Figure~\ref{fig:evolution} is necessary to reproduce the flat-bottomed shape of the EDGES signal.  We note that this results achieved with our evolving $f_{\rm esc}^{\rm X}$ could be realized if some other heating source (such as high-mass X-ray binaries) evolved similarly over this redshift range.  Note that although we do not include it here, we expect that Ly$\alpha$ heating (see \S\ref{subsec:xray_heating}) probably would not contribute meaningfully to the heating rate in our best-fitting model.  \citet{Reis2021} found that even in optimistic scenario, Ly$\alpha$ and CMB heating can add only a few 10s of K to the gas temperature by $z = 10$, while in even our low-$f_{\rm esc}^{\rm X}$ models, the gas is a few hundred K by then.    

We next ask whether the extreme UV emission required by the \textsc{very shallow-$\rho_{\rm UV}$} model, which is required in our model to reproduce the shape of EDGES is allowed by upper limits on the high-redshift contribution to the near-IR sky surface brightness (SB) from AGN.  Based off measurements of the IR-X-Ray cross-correlation~\citep{Cappelluti2013,Kashlinsky2015,MitchellWynne2016},~\citet{Windhorst2018} argued that the integrated SB at $2 \mu$m sourced by Pop III stars and AGN at $z > 7$ should be $\gtrsim 31$ mag/arcsec$^2$, or $\lesssim  0.04$ nW m$^{-2}$ sr$^{-1}$.  For a cutoff of $M_{\rm UV}^{\rm cut} = -11$, our \textsc{very shallow-$\rho_{\rm UV}$} has an integrated SB at $2\mu$m of $0.035$ nW m$^{-2}$ sr$^{-1}$ from UV emissions at $7 < z < 17$, which is close to this limit.  However, there are several reasons that this scenario likely does not violate SB constraints.  First, our model only assumes that most UV light is sourced by AGN at $z \gtrsim 10$, and integrating down to this redshift gives a SB of only $0.01$ nW m$^{-2}$ sr$^{-1}$. Second, our best-fitting model in Figure~\ref{fig:best_fit} requires only $M_{\rm UV}^{\rm cut} = -15$, for which we obtain a much more modest $0.002$ nW m$^{-2}$ sr$^{-1}$, well below the upper limit.  Lastly, since these constraints are based on the IR-X-Ray cross-correlation, it is unclear whether they would translate to the same SB limits in our model, since EDGES requires suppressed X-Ray emission from AGN relative to low-redshift expectations.  A full investigation of the IR-X-Ray cross-correlation signal in our scenario is beyond the scope of this work.  

We performed similar calculations for the integrated X-ray and radio brightnesses at $z = 0$, and compare these results to limits from Chandra~\citep[in the COSMOS field,][]{Cappelluti2017} and ARCADE~\cite{Fixsen2011}, respectively.  If we integrate the X-ray emission in our best-fit model up to $z = 10$, and redshift this to $z = 0$, we get an integrated surface brightness of $0.32$ keV cm$^{-2}$ s$^{-1}$ sr$^{-1}$ keV$^{-1}$, well below the Chandra limit of $10.91$ keV cm$^{-2}$ s$^{-1}$ sr$^{-1}$ keV$^{-1}$.  Even integrating all the way down to $M_{\rm UV} = -11$ in this case only gives $2.89$ keV cm$^{-2}$ s$^{-1}$ sr$^{-1}$ keV$^{-1}$, about a third of the observed background.  The same calculation for radio gives $258$mK at $1.4$ GHz for the fiducial model, which is marginally below the ARCADE limit of $480$ mK.  However, in this case, going to much fainter $M_{\rm UV} = -11$ increases the amplitude to $3200$ mK, well in excess of the ARCADE limit. Cutting off the integral at $z = 14$ instead of $10$ reduces these numbers to $80$ mK and $1606$ mK, respectively.  Thus, although the most extreme model in our parameter space is disallowed by the ARCADE measurement, the best-fitting model in Figure~\ref{fig:best_fit} is within the allowed parameter space.  Thus, we find that our model that best fits the EDGES measurement in amplitude and shape does not violate constraints on the $z = 0$ extragalactic background in the UV, X-ray, or radio. 

Our findings suggest that recovering the shape of the EDGES signal requires, at minimum (1) shallower redshift evolution of the UVLF than in our \textsc{shallow-$\rho_{\rm UV}$} model up to $z \approx 20$, (2) a rapid shutoff in UV emission from all sources at $z > 21.5$ and (3) some evolution in the X-ray escape fraction of high-redshift AGN (or of some other heating source). These results are qualitatively consistent with those of~\citet{Mittal2022}.  In particular, explored observationally-motivated models for the evolution in AGN abundance and radio emission properties, including evolution in their radio-loud fraction and $M_{\rm UV}^{\rm cut}$ motivated by radio observations and JWST data at lower redshifts.  We found that these models cannot reproduce the rapid saturation and flat-bottomed shape of EDGES without additional help from evolution in X-ray properties.  

\section{Implications for high-redshift AGN properties}
\label{sec:implications}

Our findings in the previous section suggest that an AGN-sourced radio background capable of explaining the EDGES signal would require fairly extreme choices of most or all the parameters involved.  These include that radio-loud AGN would have to be the dominant sources of UV emission at $z > 15$, and have very low production and/or escape of X-ray photons compared to expectations based on lower-redshift expectations.  Given these and other extreme requirements discussed above, {\it it seems unlikely that radio-loud AGN are responsible for EDGES}.  Nonetheless, here we briefly speculate about the astrophysical mechanisms that could be at play if this is the case.  

The first, most obvious implication of our best-fitting model is that the UVLF transitions from being dominated by galaxies to AGN over the redshift range $10 \lesssim z \lesssim 15$.  This is the opposite of the trend observed at $z < 10$, where the fraction of UV light produced by AGN is observed to decline with redshift~\citep{Finkelstein2022}.  However, it is also qualitatively consistent with observations suggesting that the black hole mass to stellar mass ratio increases with redshift~\citep{Ding2017,Pacucci2024,Maiolino2024} relative to the canonical relation at $z = 0$~\citep{Kormendy2013}.  A large population of obscured AGN at high redshifts would also help to explain the abundance of massive quasars at $z \gtrsim 6$, which seem to require faster growth rates than suggested by the observed AGN population~\citep{Morey2021,Eilers2021}.  Distinguishing these scenarios will require targeted spectroscopic follow-up of high-redshift objects to determine whether faint AGN are ubiquitous in the early universe.  

One mechanism that has been proposed to generate an over-abundance of accreting black holes at very high redshift is that of direct collapse~\citep{Latif2016}.  In this scenario, dissociation of H2 by the Lyman-Werner background prevents star formation in gas clouds, increasing the Jeans scale and causing gas to collapse directly into a black hole.  If direct collapse black holes (DCBHs) are ubiquitous in the early universe, they could explain the predominance of AGN required in our models.  Some recent efforts have been made to identify DCBH candidates at high redshift with JWST~\citep{Nabizadeh2024}.  Singling out this scenario will require a better understanding of the black hole mass function at high redshifts~\citep{Sicilia2022}.  

Another implication of our findings is that the production and/or escape of X-ray photons is very low at $z \gtrsim 15$, and grows (perhaps abruptly) towards lower redshifts.  Evolution in X-ray escape from AGN could be explained if nearly all AGN at very high redshift are highly obscured, by dense HI clouds and/or dust~\citep{Gilli2022,Satyavolu2023,Mazzolari2024}, which would require an increase in the obscured fraction with redshift~\citep{Vijarnwannaluk2022,Peca2023}.The abrupt evolution in $f_{\rm esc}^{\rm X}$ at $z \sim 15$ in Figure~\ref{fig:evolution} is consistent a transition from column densities of $\log(N_{\rm HI}/{\rm cm}^{-2}) \approx 25.8$ to $\approx 21.4$, more typical of the columns associated with HMXRBs~\citep{Das2017}.  If these AGN are obscured and have low X-ray escape, they would also have low LyC escape fractions, and would likely not start reionization at $z > 15$.  If escape fractions rapidly increase after this, it may explain the presence of recently-formed ionized bubbles around some UV-luminous sources near this redshift~\citep[e.g.][]{Witstok2025,Naidu2025}.  Another possibility is that most AGN at these redshift are in a unique evolutionary stage that enhances their radio production~\citep[e.g.][]{Patil2020} and/or intrinsic X-ray production~\citep{Pouliasis2024}.  

We briefly comment on the possibility that the AGN responsible for back-lighting EDGES are Little Red Dots~\citep[LRDs,][]{Matthee2024}.  While the physical nature of LRDs is debated, a number of works have suggested that they could be AGN~\citep{Ananna2024} that produce little or no X-ray emission~\citep{Ananna2024b,Sacchi2025}.  Their lack of X-ray emission may make them a possible candidate to satisfy the conditions described in the previous section.  Unfortunately, it is presently unknown whether LRDs emit significantly in the radio~\citep{Latif2025}.  If LRDs are ever observed to produce significant radio emission, it would warrant a follow-up study similar to this one to determine if they could help explain the EDGES signal.  

\section{Conclusions}
\label{sec:conc}

We have investigated the possibility that a population of faint, radio-loud AGN sourced an excess radio background that explains the anomalous depth of the EDGES 21 cm signal.  Our main findings are: 

\begin{itemize}

    \item Producing the minimum radio background that can explain the $-500$ mK depth of the EDGES signal at $z = 17$ requires that (1) most $z > 15$ UV luminous objects down to $M_{\rm UV} = -15$ or fainter are radio-loud AGN, and (2) the redshift evolution of the UVLF up to $z = 20$ is on the shallow end of what is expected based on the latest JWST measurements at $10 < z < 14$.  

    \item Assuming these objects have intrinsic X-ray luminosities comparable to those observed at lower redshifts, the escape fraction of X-ray photons must be $\lesssim 1\%$ at $z > 15$ to avoid washing out the signal via X-ray pre-heating.  This result holds regardless of how quickly the UVLF evolves because AGN are sources of both radio and X-ray photons.  

    \item Reproducing the sharp redshift evolution on either side of the EDGES absorption trough requires (1) a rapid decline in UV emission from all sources at $z > 21.5$ and (2) rapid redshift evolution in the X-ray escape fraction and/or intrinsic X-ray emission of AGN, or of some other heating source such as X-ray binaries.  Evolution in the radio emission properties of AGN and their contribution to the UV luminosity function alone cannot reproduce the rapid saturation of the EDGES signal at $z \sim 15$.  
    
\end{itemize}

These extreme requirements render our model an unlikely cosmological explanation for the unexpected depth and shape of the EDGES signal.  Our model would suggest that the UVLF transitions from being dominated by stellar sources at $z \sim 10$ to AGN at $z \sim 15$ - a cosmic time interval of just $\approx 200$ Myr (in our cosmology).  It would also require the physical conditions dictating X-ray production and/or escape to evolve.  These include the fraction of AGN that are obscured by dense HI and/or the detailed physics driving the production of radio and X-ray photons.  The requirements of our model will soon be tested by the UV, radio, and X-ray properties of high-redshift galaxies and AGN.

\paragraph{Acknowledgments}
The authors thank Sahil Hegde, Steven Furlanetto, Anson D'Aloisio, Timothy Carleton, and Jens Chluba for helpful discussions and/or comments on the draft version of this manuscript. 

\paragraph{Funding Statement}
AN acknowledges the support of the RENTU program at Arizona State University.  CC acknowledges support from the Beus Center for Cosmic Foundations at Arizona State University. RAW acknowledges support from NASA JWST Interdisciplinary Scientist grants
NAG5-12460, NNX14AN10G and 80NSSC18K0200 from GSFC. JDB acknowledges support from NSF award AST-2206766.

\paragraph{Competing Interests}
The authors are not aware of any competing interests connected to this work.  

\paragraph{Data Availability Statement}
The data underlying this article will be shared upon reasonable request to the corresponding author.

\paragraph{Ethical Standards}
The research meets all ethical guidelines, including adherence to the legal requirements of the study country.

\paragraph{Author Contributions}

AN wrote code, ran calculations, made figures, and helped draft the manuscript.  CC provided project guidance, helped check code, and contributed heavily to the drafting of the manuscript.  JDS, JB, and RW provided project guidance and feedback on the manuscript.

\printendnotes

\printbibliography 
\appendix


\end{document}